\documentclass[11pt]{article}
\usepackage{geometry}                % See geometry.pdf to learn the layout options. There are lots.
\usepackage[parfill]{parskip}
\usepackage{amsfonts,amsmath,amsthm,array,amssymb}
\newtheorem{theorem}{Theorem}

\usepackage{graphicx}
\usepackage{picinpar,graphicx}
\usepackage{graphics}
\usepackage{multicol}
\usepackage{tabularx}
\usepackage{ctable}
\usepackage{booktabs}
\usepackage{setspace}
\usepackage{epsfig}
\usepackage{graphics}
\usepackage{epstopdf}

\begin{document}
%%---------------------------------------------------------------------------%%
%%-----------------------TITLE PAGE-----------------------------------%%
%%---------------------------------------------------------------------------%%
\title{Ticks, Deer, Mice, and a Touch of Sensitivity: A Recipe for Controlling Lyme Disease}
\author{Matthew Jastresbki$^{1}$, Joan Ponce$^{2}$, \\ Daniel Burkow$^{3}$, Oyita Udiani$^{3}$, \\ Dr. Leon Arriola$^{4}$}
\date{\today}
\maketitle
\begin{center}
\footnotesize $^{1}$ Applied Mathematics, Northeastern Illinois University, Chicago, Illinois\\
\footnotesize $^{2}$ Pure Math, University of Florida, Gainsville, Florida\\
\footnotesize $^{3}$ Applied Mathematics, Arizona State University, Tempe, Arizona\\
\footnotesize $^{4}$ Pure Math, University of Wisconsin--Whitewater, Whitewater, Wisconsin\\
%\footnotesize $^{5}$ Department, School, City, State\\
%\footnotesize $^{5}$ Department, School, City, State\\
\end{center}

%%-----------------------------------------------------------------------------%%
%%-----------------------ABSTRACT--------------------------------------%%
%%-----------------------------------------------------------------------------%%
\begin{abstract} 
{\emph{Borrelia burgdorferi sensu stricto} is a bacterial spirochete prevalent in the Northeastern United States that causes Lyme disease. Lyme disease is the most common arthropod--borne disease in the United States; affecting mice, deer, humans and other mammals. The disease is spread by  \emph{Ixodes Scapularis}, a species of tick whose primary food source are deer and mice. Reducing the population of ticks feeding on both large and small mammals below some critical threshold can decrease the prevalence of Lyme disease among humans. A simplified, six--dimensional Susceptible--Infected, SI, model is used to capture the mice--deer--tick dynamics while considering the impact of varying population--specific death rates on infected population size. We analyzed the stability of the models two equilibria, the unstable disease free equilibrium and the endemic equilibrium. Static forward sensitivity analysis is conducted on the basic reproduction number and the endemic equilibrium. A dynamic approach was explored to observe change in the sensitivity of the death rates over time. These analyses were conducted to determine the efficacy of changing death rates in order to reduce prevalence of Lyme disease.
}
\end{abstract}

%%-------------------------------------------------------------------------------%%
%%-----------------------INTRODUCTION---------------------------------%%
%%-------------------------------------------------------------------------------%%

\renewcommand{\baselinestretch}{2} %for double spacing make this number 2

\section{Introduction}

Lyme disease is an epidemic in the U.S. and is caused by the bacterium, \emph{Borrelia burgdorferi Sensu Stricto}. \emph{B.\ burgdorferi} is transmitted in the Northeast and Midwest by a tick, \emph{Ixodes scapularis}, and was first discovered in the area of Lyme, Connecticut in 1977 \cite{Lyme2013wars}. The disease is heavily concentrated in the Northeast and upper Midwest of the United States; causing victims to exhibit symptoms including fever, headache, fatigue, and a characteristic skin rash called \emph{erythema migrans}. If left untreated, the infection can spread to joints, the heart, and the nervous system \cite{Lymedata}. If properly detected, Lyme disease is usually treated with antibiotics in its early stages and no further complications ensue\cite{steere1983treatment}; however, up to twenty percent of the people infected with Lyme disease exhibit symptoms that can last many years after the treatment. The chronic symptoms, referred to as Post--Treatment Lyme Disease Syndrome (PTLDS) \cite{biesiada2012Lyme}, vary but can include muscle and joint pains, cognitive defects, sleep disturbance, fatigue, seizures, and even death. The consequences of Lyme disease and PTLDS are extremely debilitating and can last for years; forcing those affected to depend on others for their most basic daily needs.

Lyme disease is the most commonly reported tick-borne illness in the United States. In 2009, the C.D.C.\ reported 38,000 documented cases nationwide; three times more than in 1991 \cite{Lyme2013wars}. In states such as Delaware, Illinois, Wisconsin, and other neighboring states, the number of cases has tripled from 2002 to 2011. Most researchers agree that the true number of infections is five to ten times higher due to undiagnosed cases. Lyme disease is classified as a zoonosis because transmission to humans from an animal reservoir is carried out by ticks that feed on both parties \cite{dunham2009live}.

Adult ticks lay their eggs in the spring after feeding on deer. During the late spring and summer, the larvae feed on small mammals such as and mice who act as natural reservoirs of the \emph{B.\ burgdorferi} spirochete \cite{bosler1984prevalence}. Nymphal ticks also feed on the mouse population with the infected nymphs passing on their infection to their new potentially susceptible host \cite{lo2004identifying}. In the final stage of its lifecycle, the adult tick will feed on a large mammal, reproduce, and then die. The deer is the preferred host of the adult \emph{I.\ Scapularis}, as these large mammals can easily survive through the winter season. 

Lyme disease can spread through fluid transfer, such as a blood transfusion, but is not sexually transmissible and is not passed on to offspring. \cite{woodrum1999investigation} No consensus exists supporting or refuting the idea that deer can transmit Lyme disease to ticks; even though the spirochete can be found on the skin and in the muscles of deer \cite{bruno2000detection}. Furthermore, the majority of studies conducted in this area claim that it is not possible. In this model, we assume deer to tick transmission is negligible, i.\ e.\ $\beta_{DT} = 0$, where $\beta_{DT}$ represents the contact rate and rate of transmission from deer to tick. The only proven method for control of the disease has been complete eradication of the deer population \cite{barbour1993biological}. This course of action was only possible due to the geographical isolation of the community. However, a reduced deer population can lead to exacerbated diseased tick density because they start to feed predominantly on mice. We do not recognize the eradication of deer to be a feasible solution to the problem of Lyme disease. Other strategies have involved attempting to control the interactions between mice and ticks in order to control the infection at its source, an approach attempted using cotton balls laced with a pesticide called permethrin in the dens of the mice. The balls made the mice unattractive as carriers for nymphal ticks \cite{barbour1993biological}. This was not a workable solution because the mouse population was so dense that the effect of the cotton balls on the total population was insignificant. Another approach is the vaccination of reservoir hosts against \emph{B.\ Burgdorferi} \cite{barbour1993biological}. This approach would be effective for a particular reservoir species, but would not be effective given the diversity of the \emph{B.\ burgdorferi} reservoir. 
	
Our work reveals a partial solution to the problem by controlling the Lyme disease vector at the host level. Our model looks to show the efficacy of modifying the death rate of one species at reducing populations of the other species. These perturbations in the death rates can be accomplished through harvesting or other prevention methods. Dual reservoirs and a single vector population are incorporated into an SI  model in six dimensions.

%While viewing this file, first use pdfLaTeX, then BibTeX, then pdfLaTeX again when you are running this for the first time or anytime you have added new references.  Otherwise they will come up as a question mark (?).

%%%%%%%%%%%%%%%%%%%%%%%%%%%%%%%%%%%%%%%%%%%%%%%%%%%%%%%%%

%%%%%%%%%%%%%%%%%%%%%%%%%%%%%%%%%%%%%%%%%%%%%%%%%%%%%%%%%

%%-------------------------------------------------------------------------------%%
%%-----------------------MODEL & METHODS---------------------------%%
%%-------------------------------------------------------------------------------%%

\section{Model \& Methods}

We construct a single--patch  Susceptible--Infected (SI) model ignoring immigration, emigration and seasonality. A removed or recovered class is not incorporated into the model as is typical in SIR models, since infected animals do not recover. Furthermore, all animal and arthropod carriers of \emph{Borellia Burgdorferi Sensu Stricto}, are asymptomatic; implying no disease-related deaths. We construct a three--tier, bi--compartmental model incorporating the vector and host species. The model yields a six-dimensional system of differential equations describing the evolution of Lyme disease in a population of deer, mice and ticks. We choose a timescale of twenty--four hours per time step, due to the time it takes the bacteria to be transmitted from a tick to a human which is thirty--six to forty--eight hours. \cite{Lymedata}

We make several assumptions for the model: we see no vertical transmission of the spirochete in ticks or any of the host species \cite{rollend2013transovarial}; it is not spread like a venereal disease, nor is it contagious; and there is no horizontal transmission of \emph{B.\ Burgdorferi} at the level of the vectors or the host species \cite{woodrum1999investigation}. Further, since studies claim that deer are a dilutant host; i.e. they are incapable of transmitting the \emph{B.\ Burgdorferi} spirochete to a feeding vector, we assume no transfer of infection from deer to tick \cite{barbour1993biological}. This assertion has been lightly contested, so assuming a transfer rate would be an interesting avenue for further study, but is beyond the scope of the current project. Since we find B. Burdorferi present on the skin of deer, we consider an infected deer class , additionally, ticks reproduce on the deer; therefore, we assume that the density of the deer population is positively correlated with the tick reproduction rate. Finally, we assume that the frequency of tick bites per unit time saturates at one. This means that out of all the contacts a given tick will have with potential hosts in a day, it will feed on only one. This allows us to simplify the contact expressions in each of our six equations, thereby simplifying the model.

The six populations described in the model are defined in Table 1:

\begin{table}[h]
\caption{Populations Involved in the Model}
\label{table:variables 1} 
\centering 
\begin{tabularx}{\textwidth}{>{} lX}
\toprule
State Variable  & Meaning \\ [0.5ex]
\toprule
$S_{D}$ & Susceptible deer population. \\ [0.5ex]
$I_{D}$ & Infected deer population. \\ [0.5ex]
$S_{M}$ & Susceptible mouse population. \\ [0.5ex]
$I_{M}$ & Infected mouse population. \\ [0.5ex]
$S_{T}$ & Susceptible tick population. \\ [0.5ex]
$I_{T}$ & Infected tick population. \\ [0.5ex]
\bottomrule
\end{tabularx}
\end{table}
\newpage

In the following table, (Table \ref{table:variables}), population densities are per square kilometer. The parameter values that we use have been acquired from census data, the CDC, hunting reports, and also from various studies conducted in the field. 
%\newpage

\begin{table}[!h]\caption{Definition of the variables in the modeling 
framework}
\label{table:variables 2} 
\centering 
\begin{tabularx}{\textwidth}{>{} l|X|c|c|}
\toprule
Parameter & Meaning & Units & Value \\ [0.5ex]
\toprule
$D$ & Total population of deer & Individuals & 300\\ [0.5ex]
$M$ &  Total population of mice & Individuals & 3500\\[0.5 ex]
$T$ &  Total population of ticks & Individuals & 1530000\\[0.5 ex]
$S_{D}$ & Susceptible deer & Individuals & 216\\[0.5ex]
$I_{D}$ & Infected deer & Individuals & 84\\[0.5ex]
$S_{M}$ & Susceptible mice  & Individuals & 1050\\ [0.5ex]	
$I_{M}$& Infected mice & Individuals & 2450\\ [0.5ex]
$S_{T}$ & Susceptible tick & Individuals & 979200\\ [0.5ex]	
$I_{T}$& Infected Ticks & Individuals & 550800\\ [0.5ex]
$\Lambda_{D}$ & Birth/recruitment rate of deer & Individuals/time     & $0.147945205 \pm ?$\footnotemark[1]\\[0.5 ex] 
$\Lambda_{M}$ & Birth/recruitment rate of mice & Individuals/time      & $5.4*10^-2 \pm ?$\footnotemark[1]\\[0.5ex]
$\Lambda_{T}$ & Birth/recruitment rate of tick & Individuals/time       & $2.6*10^-2 \pm ?$\footnotemark[1]\\[0.5ex]
$\beta_{TD}$ & Transmission rate from ticks to susceptible deer	& 1/time	& $1.3699*10^-3 \pm ?$\footnotemark[1]\\[0.5ex]
$\beta_{TM}$  & Transmission rate from ticks to susceptible mice       & 1/time     & $2.739*10^-3 \pm ?$\footnotemark[1]\\[0.5 ex]
$\beta_{MT}$& Transmission rate from mice to susceptible ticks       & 1/time     & $1.863*10^-3 \pm ?$\footnotemark[1]\\[0.5 ex]
$\mu_{D}$  & Natural death rate for deer       & 1/time     & $4.9*10^-4 \pm ?$\footnotemark[1]\\[0.5 ex]
$\mu_{M}$ & Natural death rate for mice       & 1/time     & $2.7*10^-3 \pm ?$\footnotemark[1]\\[0.5 ex]
$\mu_{T}$ & Natural death rate for ticks       & 1/time     & $9.6*10^-3 \pm ?$\footnotemark[1]\\[0.5 ex]
\bottomrule
\end{tabularx}
\end{table}
\footnotetext[1]{These parameters have been taken from various sources in literature as well as CDC reports. No two sources have had matching parameter values. Therefore, values have been averaged based on all reported values. See Appendix for sources and justification.}
\newpage
\newpage
The assumptions and definitions lead to the following model on the dynamics of deer--mice--ticks:

\begin{figure}
\begin{center}
\includegraphics[width=15cm,height=10cm]{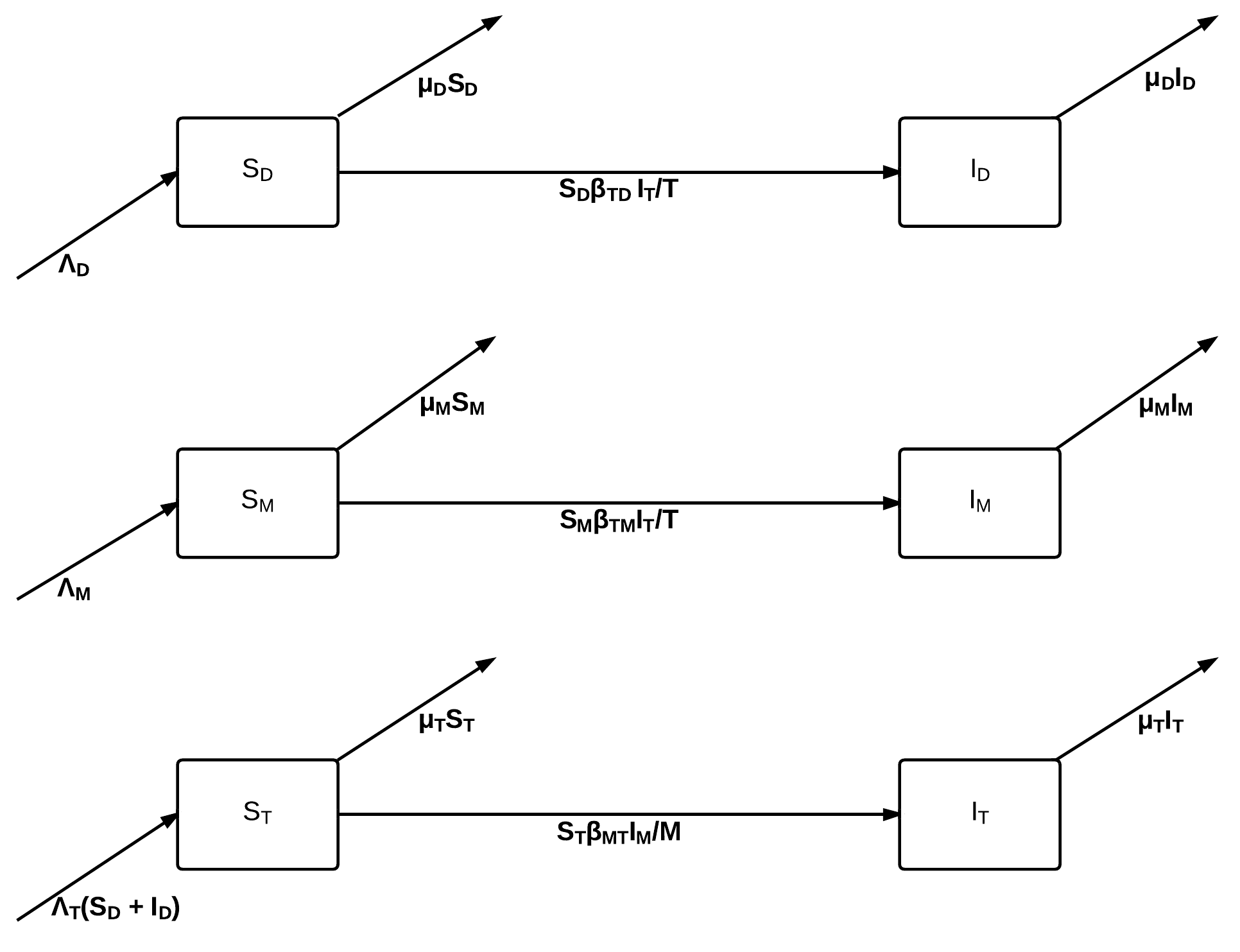}
\end{center}
\caption{\label{fig:1}Compartmental Model}
\end{figure}

\begin{equation}
\label{model1}
{} \quad
\begin{cases}
\dot{S}_{D} = \Lambda _{D}-\beta_{TD}\dfrac{I_{T}}{T}S_{D}-\mu _{D}S_{D}\vspace{1.5mm},\\

\dot{I}_{D} = \beta_{TD}\dfrac{I_{T}}{T}S_{D}-\mu _{D}I_{D}\vspace{1.5mm},\\

\dot{S}_{M} = \Lambda _{M}-\beta _{TM}\dfrac {I_{T}} {T}S_{M}-\mu _{M}S_{M} \vspace{1.5mm}\\

\dot{I}_{M} = \beta _{TM}\dfrac {I_{T}} {T}S_{M}-\mu _{M}I_{M} \vspace{1.5mm}\\

\dot{S}_{T} = \Lambda _{T}D-\beta _{MT}S_{T}\dfrac {I_{M}} {M}-\mu_{T}S_{T} \vspace{1.5mm}\\

\dot{I}_{T} = \beta _{MT}S_{T}\dfrac {I_{M}} {M}-\mu_{T}I_{T} \vspace{1.5mm}\\
\end{cases}
\end{equation}

Where:
$D=S_{D}+I_{D}$, 
$T=S_{T}+I_{T}$ and    
$M=S_{M}+I_{M}$
%\newpage

In the equations in \eqref{model1}, where $i \neq j = D, M, T$, the susceptible population that gets infected moves into the infected class with a per capita rate of $\beta _{ij}\dfrac {I_{i}} {i}$. $\beta _{ij}$ is the contact rate and rate of transmission of the spirochete from population $i$ to population $j$. The $\dfrac {I_{i}} {i}$ represents the proportion of infected members of population $i$ that a given individual in population $j$ encounters. The susceptible and infected individuals also leave their respective classes through death according to the terms $\mu _{i}S_{i}$ and $\mu _{i}I_{i}$ respectively. The terms $\mu_{i}$ represent the per capita death rate of population $i$ in each time step. New individuals are recruited into the susceptible population via $\Lambda_{i}$, which represents the number of units of population $i$ that are born in each time step. Note that $\Lambda_{D}$ and $\Lambda_{M}$ are constant while $\Lambda_{T}$ is dependent on the density of deer. \\

\begin{figure}[h!]
\begin{center}
\includegraphics[width=15cm,height=10cm]{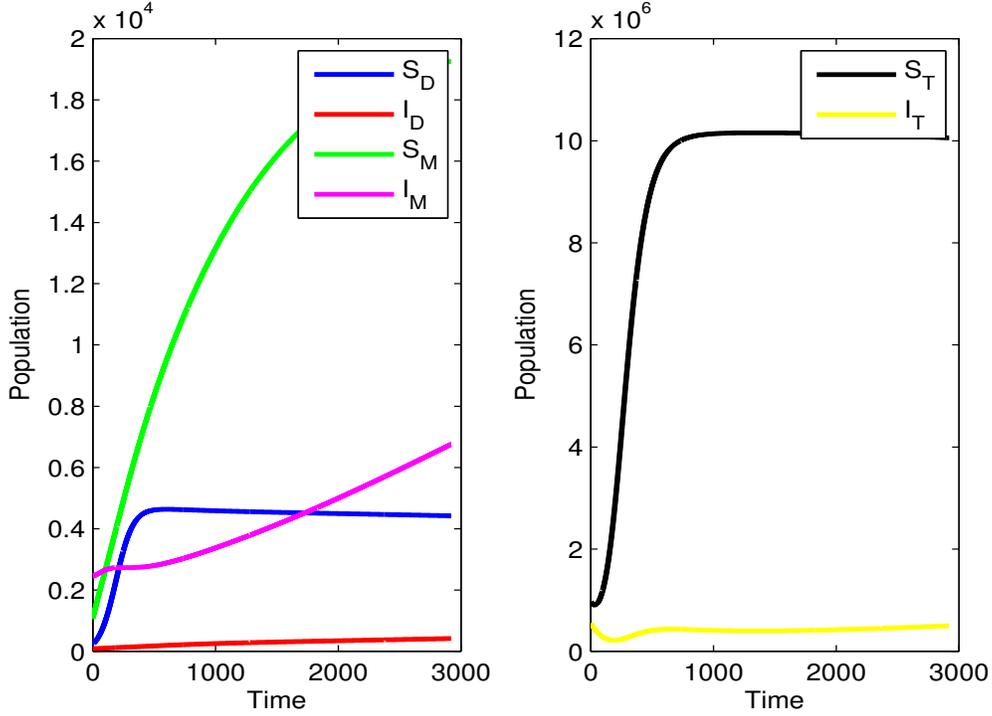}
\end{center}
\caption{\label{fig:2}Population Dynamics}
\end{figure}

The simulation \eqref{fig:2} was run over a period of 3000 days to observe the long--term and transient behavior of the model incorporating the initial conditions listed in Table \ref{table:variables}. 

%%%%%%%%%%%%%%%%%%%%%%%%%%%%%%%%%%%%%%%%%%%%%%%%%%%%%%%%%%%%%%%%%%%%%%%%%%%%%%%%%%%%%%%%%%%%%%%%%

%%--------------------------------------------------------------------------------------------------------------------%%
%%----------DISEASE-FREE AND ENDEMIC EQUILIBRIA AND THEIR STABILITY------------%%
%%--------------------------------------------------------------------------------------------------------------------%%

\section{Disease-Free and Endemic Equilibria and Their Stability}

\subsection{Disease-Free Equilibria and Stability }

The Disease--Free equilibrium (DFE) occurs when the pathogen has suffered extinction and every individual of the population is susceptible. Therefore, in the system \eqref{model1} we compute the disease--free equilibrium by setting $\text{I}_D=0$, $\text{I}_M=0$ and $\text{I}_T=0$.

The system \eqref{model1} has a disease free equilibrium denoted by $E_0$, where
$$ E_0=[S_D\to\frac{\Lambda _D}{\mu _D},\text{I}_D\to 0, S_M\to \frac{\Lambda _M}{\mu
   _M},\text{I}_M\to 0,S_T\to \frac{\Lambda _D \Lambda _T}{\mu _D \mu _T},\text{I}_T\to 0]$$

The basic Reproduction number, $\mathcal R_0$, represents the number of new infections one case generates on average over the course of its infectious period.

In order to compute $\mathcal R_0$, we use the next generation matrix \cite{diekmann1990definition} and obtain the following:

    \[  X = \begin{pmatrix}
               \text{I}_D\\
              \text{I}_M \\
              \text{I}_T\\
              \end{pmatrix}
                \qquad\text{,    }\qquad
        Y = \begin{pmatrix}
               \text{S}_D \\
               \text{S}_M \\
                \text{S}_T \\
                \end{pmatrix}
    \]

Where $X$ is the vector of infected classes and Y the vector of uninfected classes

The original system of equations can be rewritten in the following generalized form:

$\dfrac {\partial X} {\partial t}=\mathcal F(X,Y) -\mathcal V(X,Y)$

    \[  \mathcal F = \begin{pmatrix}
 \frac{\text{S}_D \text{I}_T  \beta _{\text{TM}}}{\text{I}_T+\text{S}_T} \\
 \frac{\text{I}_T S_M \beta _{\text{TM}}}{\text{I}_T+S_T}  \\
 \frac{\text{I}_M \beta _{\text{MT}} S_T}{\text{I}_M+S_M} \\
              \end{pmatrix}
                \qquad\text{  ,  }\qquad
        \mathcal V= \begin{pmatrix}
\mu _{D}S_D \\
\mu _{M}I_{M}  \\
\mu _{T}I_{T} \\
                \end{pmatrix}
    \]

Let $F=J(\mathcal F)$ and $V=J(\mathcal V)$, where $J$ denotes the Jacobian. Next, compute the eigenvalues of the associated matrix $FV^{-1}$. The reproduction number  will be the largest eigenvalue of the Jacobian, thus:\\
\begin{equation}
\mathcal R_0=\sqrt {\dfrac {\beta _{TM}\beta _{MT}} {\mu _{M}\mu _{T}}}
\end{equation}

\begin{theorem} If $\mathcal R_0<1$, then the disease free equilibrium, $ E_0$, is locally asymptotically stable for the system \eqref{model1}.  If $\mathcal R_0>1$, then $ E_0$ is unstable. \cite{van2002reproduction}

\end{theorem}

If $\mu _{M}\mu _{T} \leq \beta _{TM}\beta _{MT}$, then $\mathcal R_0>1$, thus the DFE will be unstable.

\subsection{Endemic Equilibria and Stability}
   
The Endemic Equilibrium of the model is obtained by considering the infectious classes to be greater than zero. The following Endemic Equilibrium is obtained:

$$E^*=(S^{*}_{D}, I^{*}_{D}, S^{*}_{M}, I^{*}_{M},S^{*}_{T}, I^{*}_{T}),$$ Where, $$S^*_D=\frac{\Lambda _D \beta _{\text{TM}} \left(\beta _{\text{MT}}+\mu _T\right)}{\mu _T \left(\mu _D \beta
   _{\text{TM}}-\mu _M \beta _{\text{TD}}\right)+\beta _{\text{MT}} \beta _{\text{TM}} \left(\mu
   _D+\beta _{\text{TD}}\right)},$$ $$I^*_D= \frac{\Lambda _D \beta _{\text{TD}} \left(\beta _{\text{MT}} \beta _{\text{TM}}-\mu _M \mu
   _T\right)}{\mu _D \left(\mu _T \left(\mu _D \beta _{\text{TM}}-\mu _M \beta
   _{\text{TD}}\right)+\beta _{\text{MT}} \beta _{\text{TM}} \left(\mu _D+\beta
   _{\text{TD}}\right)\right)},$$ $$S^*_M=\frac{\Lambda _M \left(\beta _{\text{MT}}+\mu _T\right)}{\beta
   _{\text{MT}} \left(\mu _M+\beta _{\text{TM}}\right)},\text{I*}_M= \frac{\Lambda _M \left(\beta
   _{\text{MT}} \beta _{\text{TM}}-\mu _M \mu _T\right)}{\mu _M \beta _{\text{MT}}
   \left(\mu _M+\beta _{\text{TM}}\right)},$$ $${S^*}_T= \frac{\Lambda _D \Lambda _T \left(\mu
   _M+\beta _{\text{TM}}\right)}{\mu _D \beta _{\text{TM}} \left(\beta _{\text{MT}}+\mu
   _T\right)},\text{I*}_T= \frac{\Lambda _D \Lambda _T \left(\beta _{\text{MT}}
   \beta _{\text{TM}}-\mu _M \mu _T\right)}{\mu _D \mu _T \beta _{\text{TM}}
   \left(\beta _{\text{MT}}+\mu _T\right)}.$$
  
 It is important to note that the reproductive number is present in the numerator of the infected terms of the endemic equilibrium, thus the infected populations are higher if the reproductive number is larger. The reproductive number, as well as the endemic equilibria depend on transmission rates between mice and ticks as well as the death rates of both populations. 

\begin{theorem} 
The system \eqref{model1} has a unique endemic equilibrium, $E^*=(S^{*}_{D}, I^{*}_{D}, S^{*}_{M}, I^{*}_{M},S^{*}_{T}, I^{*}_{T})$, iff $\mathcal R_0>1$ and $\mathcal R_0 > \mathcal R_1$, where $\mathcal R_1 = \frac{\mu_M \beta_{TD} - \mu_D\beta_{TM}}{(\mu_D+\beta_{TD} \mu_M)}$. The endemic equilibrium $E^*$ is locally asymptotically stable whenever it exists.
\end{theorem}

In order to verify stability we compute the Jacobian evaluated at the endemic equilibrium:
\begin{equation}
J|_{E^*}=
\left(
\begin{array}{cccccc}
 -\frac{\text{I}_T \beta _{\text{TD}}}{\text{I}_T+S_T}-\mu _D & 0 & 0 & 0 & \frac{\text{I}_T
   S_D \beta _{\text{TD}}}{\left(\text{I}_T+S_T\right){}^2} & -\frac{S_D S_T \beta
   _{\text{TD}}}{\left(\text{I}_T+S_T\right){}^2} \\
 \frac{\text{I}_T \beta _{\text{TD}}}{\text{I}_T+S_T} & -\mu _D & 0 & 0 & -\frac{\text{I}_T S_D
   \beta _{\text{TD}}}{\left(\text{I}_T+S_T\right){}^2} & \frac{S_D S_T \beta
   _{\text{TD}}}{\left(\text{I}_T+S_T\right){}^2} \\
 0 & 0 & -\frac{\text{I}_T \beta _{\text{TM}}}{\text{I}_T+S_T}-\mu _M & 0 & \frac{\text{I}_T
   S_M \beta _{\text{TM}}}{\left(\text{I}_T+S_T\right){}^2} & -\frac{S_M S_T \beta
   _{\text{TM}}}{\left(\text{I}_T+S_T\right){}^2} \\
 0 & 0 & \frac{\text{I}_T \beta _{\text{TM}}}{\text{I}_T+S_T} & -\mu _M & -\frac{\text{I}_T S_M
   \beta _{\text{TM}}}{\left(\text{I}_T+S_T\right){}^2} & \frac{S_M S_T \beta
   _{\text{TM}}}{\left(\text{I}_T+S_T\right){}^2} \\
 \Lambda _T & \Lambda _T & \frac{\text{I}_M S_T \beta
   _{\text{MT}}}{\left(\text{I}_M+S_M\right){}^2} & -\frac{S_M S_T \beta
   _{\text{MT}}}{\left(\text{I}_M+S_M\right){}^2} & -\frac{\text{I}_M \beta
   _{\text{MT}}}{\text{I}_M+S_M}-\mu _T & 0 \\
 0 & 0 & -\frac{\text{I}_M S_T \beta _{\text{MT}}}{\left(\text{I}_M+S_M\right){}^2} & \frac{S_M
   S_T \beta _{\text{MT}}}{\left(\text{I}_M+S_M\right){}^2} & \frac{\text{I}_M \beta
   _{\text{MT}}}{\text{I}_M+S_M} & -\mu _T \\
\end{array}
\right)
\end{equation}
Which yields the following eigenvalues:

\begin{equation}
\begin{array}{cccc}
\lambda_{1}=-\mu_{D}\\
\lambda_{2}=-\mu_{T}\\
\lambda_{3}=-\mu_{M}\\
\lambda_{4}=\frac{\mu _T \left(\mu _M \beta _{\text{TD}}-\mu _D \beta
   _{\text{TM}}\right)-\beta _{\text{MT}} \beta _{\text{TM}} \left(\mu _D+\beta
   _{\text{TD}}\right)}{\beta _{\text{TM}} \left(\beta _{\text{MT}}+\mu _T\right)}\\
\end{array}
\end{equation}
The first three eigenvalues are negative and $\lambda_{4}<0$ when $\mathcal R_0 > \mathcal R_1$. $\lambda_{5,6}$ are the solutions the quadratic equation of the form $A\lambda_{5,6}^2+B\lambda_{5,6}+C=0$, where:
\begin{center}
 $A = \left(\mu _M+\beta _{\text{TM}}\right) \left(\beta _{\text{MT}}+\mu _T\right)$,\\
 $B = \beta _{\text{MT}} \left(\left(\mu _M+\beta _{\text{TM}}\right){}^2+2 \mu _T \beta
   _{\text{TM}}\right)+\beta _{\text{MT}}^2 \beta _{\text{TM}}+\mu _T^2 \beta _{\text{TM}})$\\
$C = \left(\mu _M+\beta _{\text{TM}}\right) \left(\beta _{\text{MT}}+\mu _T\right) \left(\beta
   _{\text{MT}} \beta _{\text{TM}}-\mu _M \mu _T\right)$

 \end{center}

It is clear that $A$ and $B$ are always greater than zero. When $\mathcal R_{0} > 1$, the following condition holds $\mu_{M}\mu_{T}\leq\beta_{TM}\beta_{MT}$.  It follows that $C$ will always be greater than or equal to zero while $\mathcal R_0>1$. Therefore, the square root of the discriminant will be between zero and $B$. Thus, the two eigenvalues of the form $\dfrac{-B \pm \sqrt{B^2 - 4AC}}{2A}$ are less than zero when $R_{0} > 1$ and the equilibrium is locally stable. 

%%%%%%%%%%%%%%%%%%%%%%%%%%%%%%%%%%%%%%%%%%%%%%%%%%%
%%------------------------------------------------------------------------%%
%%------------------SENSITIVITY ANALYSIS---------------------%%
%%------------------------------------------------------------------------%%

\section{Sensitivity Analysis}

The basic tenet of sensitivity analysis is that perturbations to the input parameters
of a model produce perturbations in the output. Sensitivity analysis quantifies these
uncertainties. The quantification is defined as the ratio of a 1 percent change
change in the parameter produces what percent change in the output.
In our case, the quantities of 
interest are both static and dynamic. Specifically, the reproduction number and the equilibrium points are static in nature, whereas the dynamic model consisting 
of the ODEs given in equations \eqref{model1} is temporal. 
Sensitivity of the reproduction number describes, via each relevant parameter,
how many secondary infections are incurred given a single infection
in a completely susceptible population. Sensitivity of the equilibrium points
describes how the long--term solutions are affected by changes in the defining 
parameters. Lastly, sensitivity of the ODE model describes the transient sensitivity.
With time dependent models, it is possible for certain parameters to exchange
relative importance. For example, parameter $p_1$ may be more important
then parameter $p_2$ up to some crossover time $t_c$. After $t_c$, parameter $p_2$
is more important than $p_1$.

Consider a generic model, as shown here, called the forward problem. It takes nominal input parameters, such as $\mu_{\text{D}}$, etc.., which we will refer to as $p$, and generates a solution $u$.
\begin{figure}[htb]
\begin{center}
\includegraphics[width=5.0truein, height=1.5truein]{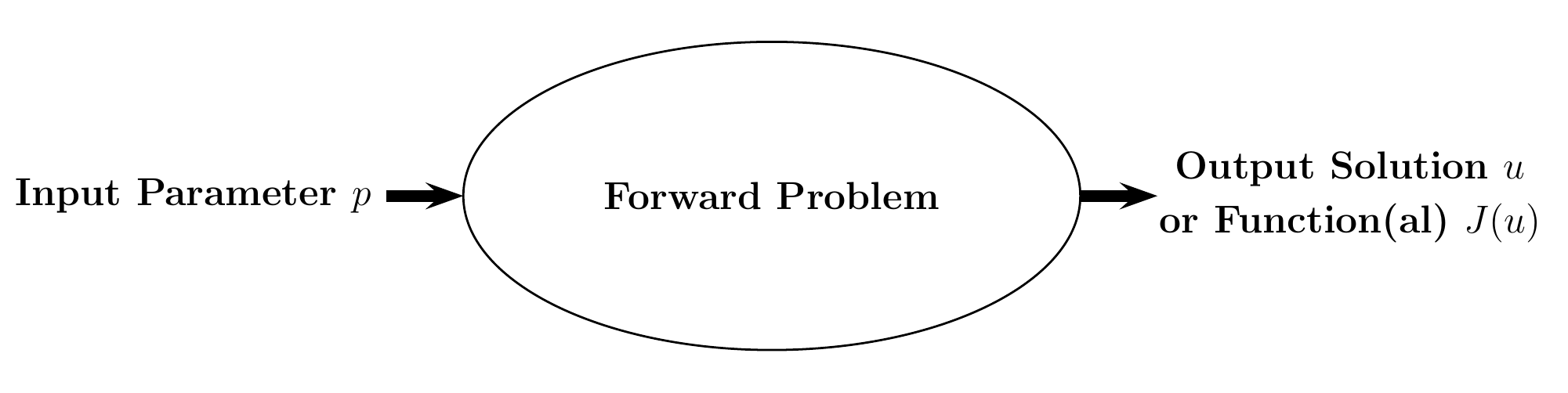}
\caption{Forward problem consists of generic input(s) $p$ and corresponding
output(s) $u$.}
\end{center}
\end{figure}

Forward sensitivity analysis introduces perturbations
to the input parameters, via $\delta p$ and quantifies the
subsequent perturbations to the output solution via $\delta u$.
\begin{figure}[htb]
\begin{center}
\includegraphics[width=5.0truein, height=1.5truein]{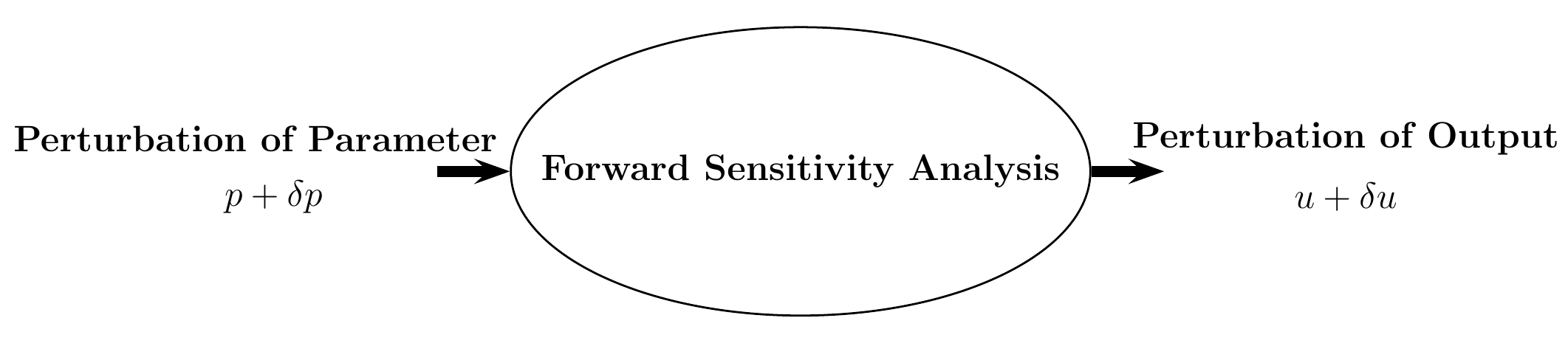}
\caption{Forward sensitivity quantifies how perturbations $\delta p$
to the input parameter $p$ produces perturbations of the output $\delta u$.}
\end{center}
\end{figure}

In order to quantify the concept of sensitivity, we define the normalized
indices.
%\begin{definition}%[Normalized Sensitivity Index]
The \textnormal{\bf normalized sensitivity index}\footnote{Some authors refer to what we call SI as elasticity. This terminology originated
from the field of economics.} 
%\textnormal{({\bf SI})}
is defined to be the limit of the ratio \cite{arriola2009sensitivity}
    \[ SI(u; p) := \lim_{\delta p \to 0}
       \displaystyle\frac{\left( \displaystyle\frac{\delta u}{u}\right)}
       {\left( \displaystyle\frac{\delta p}{p}\right)} =
       \displaystyle\frac{p}{u}
       \displaystyle\frac{\partial u}{\partial p}\qquad u \neq 0. \]
%\end{definition}
These indices essentially gives the percent change in output for a given
percent change to the input parameter. 

\subsection{Sensitivity of the Reproductive Number}

The basic reproductive number  measures the number of secondary infections created by a single infected tick/mouse in a completely susceptible population. For our model, this metric  depends on the following variables: $\mu _{M}$, $\beta _{TM}$, $\beta _{MT}$ and $\mu _{T}$.  The parameters that can feasibly be modified are the death rates of the mice, and possibly the tick population; though the latter would be considerably more difficult. The sensitivity analysis on the reproductive number is done to understand the impact on the individual populations of perturbing the death rate of the mice and the death rate of the tick population on the spread of the disease. Calculating the normalized sensitivity index of $\mathcal R_{0}$ with respect to $\mu _{M}$ and $\mu _{T}$ indicate that:
\begin{equation*}
\begin{aligned}
SI(\mathcal R_{0}, \mu_{M})&:=SI(\mathcal R_{0}, \mu_{T})\\
&=\dfrac{\mu _{M}}{ \mathcal R_{0}} \dfrac {\partial \mathcal R_{0}} {\partial \mu _{M}} \\
 &=\dfrac{\mu _{T}}{ \mathcal R_{0}}\dfrac {\partial \mathcal R_{0}} {\partial \mu _{T}}\\
&=-\dfrac{1}{2}
 \end{aligned}
\end{equation*}

Thus, if we increase the death rate of the mice (or Ticks) by one percent, the reproductive number will decrease by 0.5\%. Similar calculations show:
\begin{equation*}
\begin{aligned}
SI(\mathcal R_{0}, \beta_{TM})&:=SI(\mathcal R_{0}, \beta_{MT})\\
&=\dfrac{\beta _{TM}}{ \mathcal R_{0}} \dfrac {\partial \mathcal R_{0}} {\partial \beta _{TM}} \\
 &=\dfrac{\beta _{MT}}{ \mathcal R_{0}}\dfrac {\partial \mathcal R_{0}} {\partial \beta _{MT}}\\
&=\dfrac{1}{2}
 \end{aligned}
\end{equation*}
Thus decreasing the transmission rates between Ticks and Mice (and vice-versa),  by one percent, will decrease the reproductive number by 0.5\%. 

\subsection{Sensitivity of the Endemic Equilibrium}

We explored the sensitivities of endemic equilibrium  infectious hosts/vector densities to perturbations in death rates of each class.  This was done in order to simulate and observe the effects of introducing preventative measures to one of the host populations. 

We performed forward sensitivity analysis on the forward problem at the endemic equilibrium in order to acquire normalized sensitivity conditions with respect to each population parameter. We calculated the normalized sensitivity conditions of the species--specific death rates with respect to the populations according to the process previously described:
%\newpage

\subsubsection{Normalized Sensitivity Conditions with respect to $\mu_{D}$}

\bigskip
\begin{center}
\begin{equation*}
\begin{aligned}
SI(I_{D};\mu_{D}) &= -{\frac {{\beta_{MT}}\,{\beta_{TD}}\,{\beta_{TM}}+2\,{\beta_{TM}}\,{\mu_{D}}\,{\it \beta_{MT}}-{\beta_{TD}}\,{\mu_{M}}\,{\mu_{T}}+2\,{\beta_{TM}}\,{\mu_{D}}\,{\mu_{T}}}{{
\beta_{MT}}\,{\beta_{TD}}\,{\beta_{TM}}+{\beta_{TM}}\,{\mu_{D}}\,{\beta_{MT}}-{\beta_{TD}
}\,{\mu_{M}}\,{\mu_{T}}+{\beta_{TM}}\,{\mu_{D}}\,{\mu_{T}}}}\\
&= - \frac{R_0(\beta_{TD}+2\mu_D)\mu_M - (\beta_{TD}\mu_M + 2\beta_{TM}\mu_D)}{R_0(\beta_{TD}+\mu_D)\mu_M - (\beta_{TD}\mu_M + \beta_{TM}\mu_D)}\\
& = - \frac{\beta_{TD}\mu_M(R_0 - 1) + 2R_0 (\mu_M- \beta_{TM}) \mu_D}{\beta_{TD}\mu_M(R_0 - 1) + R_0 (\mu_M- \beta_{TM}) \mu_D}
 \end{aligned}
\end{equation*}
\begin{equation*}
\begin{aligned}
 SI(I_{M};\mu_{D}) &= 0 \\
SI(I_{T};\mu_{D}) &= -1
 \end{aligned}
\end{equation*}

\end{center}
From the equations, we can see that increasing Deer death rates always has a negative impact on $I_T$ densities. Interestingly, the this parameter does appear not dampen infected mice population (i.e. sensitivity index is zero). We note in passing that this is likely a consequence of our first order approximation of $\frac{\partial I_{M}}{\partial \mu_D}$. Finally, we observe that the sensitivity index of the infected Deer class may be positive or negative depending on the relative magnitude of $R_0$. Thus we conlcude that $\mu)_D$ has its greatest effect on $I_D$ but it always has a negative effect on $I_T$

\subsubsection {Normalized Sensitivity Conditions with respect to $\mu_{M}$}
\bigskip

\begin{equation*}
\begin{aligned}
SI(I_{D};\mu_{M}) &= -{\frac {{\mu_{T}}\, \left( {\beta_{MT}}+{\mu_{T}} \right) {\beta_{TM}}\,{\mu_{M}}\,{\mu_{D}}}{ \left( {\beta_{MT}}\,{\beta_{TD}}\,{\beta_{TM}}+{\beta_{MT}}\,{
\beta_{TM}}\,{\mu_{D}}-{\beta_{TD}}\,{\mu_{M}}\,{\mu_{T}}+{\beta_{TM}}\,{\mu_{D}}\,
{\mu_{T}} \right)  \left( {\beta_{TM}}\,{\beta_{MT}}-{\mu_{M}}\,{\mu_{T}}
 \right) }}\\
&= - \frac{(\beta_{MT} + \mu_T) \beta_{TM} \mu_D \mu_M} {\left(  R_0 (\beta_{TD} + \mu_M)\mu_M - (\beta_{TD}\mu_M - \beta{TM}\mu_D) \right) (R_0 - 1)}
\end{aligned}
\end{equation*}

\bigskip

\begin{equation*}
\begin{aligned}
SI(I_{M};\mu_{M}) & = -{\frac {{\beta_{MT}}\,{{\beta_{TM}}}^{2}+2\,{\beta_{MT}}\,{\beta_{TM}}\,{\mu_{M}}-
{\mu_{T}}\,{{\mu_{M}}}^{2}}{ \left( {\beta_{TM}}+{\mu_{M}} \right)  \left( R_0- 1 \right) \mu_M\mu_T }}\\
& = - \frac{R_0(\beta_{TM}+2\mu_M) - \mu_M}{(\beta_{TM}+\mu_M)(R_0-1)}
\end{aligned}
\end{equation*}

\bigskip
 
\begin{equation*}
\begin{aligned}
SI(I_{T};\mu_{M}) & = -{\frac {{\mu_{M}}\,{\mu_{T}}}{{\beta_{TM}}\,{\beta_{MT}}-{\mu_{M}}\,{\mu_{T}}}}\\
&= -\frac{1}{R_0 - 1}
\end{aligned}
\end{equation*}

%\newpage

From the equations, we observe that the sensitivity of the infected Deer class can also vary depending on the disease dynamics near equilibrium. If $R_0 >> 1$ (and the endemic equilibrium exits), then increasing mice death rates has a negative impact on the infected Deer class. Moreover, the impact on the infected mice and Tick populations is also negative provided $R_0 > 1$. However, when $R_0 < 1$, we see an increase in the infected tick population; note that an increase in the population does not imply an increasing population with respect to time.

\subsection{Time Sensitivity of the ODEs}

The results of our local sensitivity analysis at the endemic state revealed that potential changes in ecological parameters governing the reproduction number may change the sensitivity of perturbations in host death rates on infectious classes. This also hints at possible changes in sensitivity based on intrinsic host-vector dynamics as the disease evolves over time. Sensitivity conditions are not always time dependent, but temporal sensitivity was observed in the response measures of the SARS epidemic in China \cite{chowell2004model}. This was significant because it showed how the two response measures switched effectiveness at a certain point in the epidemic. 
We explored numerically the temporal sensitivities of these death rates by integrating  the associated forward sensitivity equations for our model. These were computed by considering the partial derivatives of our model with respect to the focal parameters: 
Differentiating the forward equations given in \eqref{model1} wrt.\ 

\[
\frac{\partial S_D}{\partial{\mu_D}}  =  -\frac{\beta_{TD}}{T^2} \bigg( (I_T \frac{\partial S_D}{\partial{\mu_D}} + S_D \frac{\partial I_T}{\partial{\mu_D}} )T - I_TS_D (\frac{\partial S_T}{\partial{\mu_D}} + \frac{\partial I_T}{\partial{\mu_D}})\bigg)
 - \mu_D \frac{\partial S_D}{\partial {\mu_D}} - S_D \]

\[ \frac{\partial I_D}{\partial {\mu_D}}  =  \frac{\beta_{TD}}{T^2} \bigg( (I_T \frac{\partial S_D}{\partial{\mu_D}} + S_D \frac{\partial I_T}{\partial{\mu_D}} )T - I_TS_D (\frac{\partial S_T}{\partial{\mu_D}} + \frac{\partial I_T}{\partial{\mu_D}})\bigg)
 - \mu_D \frac{\partial I_D}{\partial {\mu_D}} - I_D \]

\[ \frac{\partial S_M}{\partial {\mu_D}}  =  -\frac{\beta_{TM}}{T^2} \bigg( (I_T \frac{\partial S_M}{\partial{\mu_D}} + S_M \frac{\partial I_T}{\partial{\mu_D}} )T - I_TS_M (\frac{\partial S_T}{\partial{\mu_D}} + \frac{\partial I_T}{\partial{\mu_D}})\bigg)
 - \mu_M \frac{\partial S_M}{\partial {\mu_D}} \]

\[ \frac{\partial I_M}{\partial {\mu_D}}  = \frac{\beta_{TM}}{T^2} \bigg( (I_T \frac{\partial S_M}{\partial{\mu_D}} + S_M \frac{\partial I_T}{\partial{\mu_D}} )T - I_TS_M (\frac{\partial S_T}{\partial{\mu_D}} + \frac{\partial I_T}{\partial{\mu_D}})\bigg)
 - \mu_M \frac{\partial I_M}{\partial {\mu_D}}  \]

\[ \frac{\partial S_T}{\partial {\mu_D}}  = \Lambda_T \bigg(\frac{\partial S_D}{\partial {\mu_D}} + \frac{\partial I_D}{\partial {\mu_D}}\bigg) - \frac{\beta_{MT}}{M^2} 
\bigg( (S_T \frac{\partial I_M}{\partial{\mu_D}} + I_M \frac{\partial S_T}{\partial{\mu_D}} )M - S_TI_M (\frac{\partial S_M}{\partial{\mu_D}} + \frac{\partial I_M}{\partial{\mu_D}})\bigg) - \mu_T \frac{\partial S_T}{\partial {\mu_D}} \]

\[ \frac{\partial I_T}{\partial {\mu_D}}  =  \frac{\beta_{MT}}{M^2} \bigg( (S_T \frac{\partial I_M}{\partial{\mu_D}} + I_M \frac{\partial S_T}{\partial{\mu_D}} )M - S_TI_M (\frac{\partial S_M}{\partial{\mu_D}} + \frac{\partial I_M}{\partial{\mu_D}})\bigg) - \mu_T \frac{\partial I_T}{\partial {\mu_D}} \]

We examined how changes in local stability of the endemic equilibria affected the sensitivity index and observed crossover periods (i.e. where Deer death rates become  more/less important on infectious classes relative to mice death rates.)  

Our simulations ran over a time period of three thousand days, showed dynamical changes in sensitivity. The graphs below display our results. The axes display the magnitude of the normalized sensitivity indicies over time in days. Each line represents a population denoted by a specific color. Both groups display the population dynamics and the sensitivities of the populations with respect to the death rates of deer and mice respectively. For the first set, we chose parameters such that the endemic equilibrium was stable. In the second set of graphs, we chose parameters such that $R_0 < 1$ using the values listed in \ref{table:variables}. 

The graph shows that the population of infected mice gets more sensitive to perturbations in the deer death rate as time goes by. Conversely,  the susceptible mouse population gets more negatively sensitive to the death rate of deer with the passage of time. An expected result, given the dependence of \emph{I.\ Scapularis} on the deer population as reproductive hosts, the susceptible tick population gets increasingly negatively sensitive to the deer death rate. 

An interesting result of the dynamic forward sensitivity analysis with respect to the death rate of the mouse population is the complete lack of affect it has on the deer population over the entire span of three thousand days. Unsurprisingly, the death rate of the mice has an immediate, increasingly negative effect on the infected  tick population. Halfway through the simulation, however, the magnitudes of both the susceptible and infected tick sensitivity change direction in a parabolic fashion, and approach an equilibrium at zero. 
\newpage

\begin{figure}[hc!]
\begin{center}

\includegraphics[width=10cm,height=5cm]{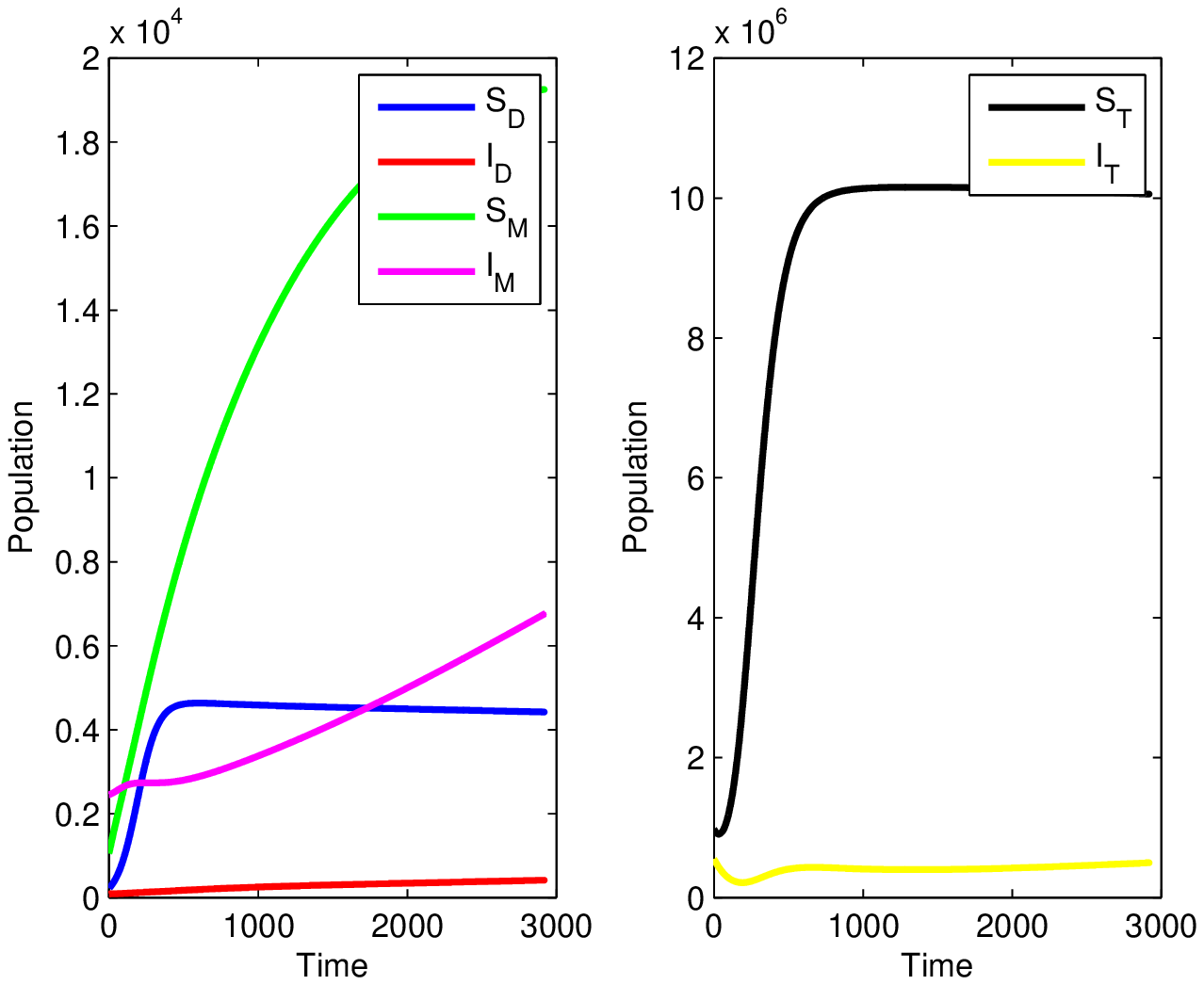}
\includegraphics[width=10cm,height=5cm]{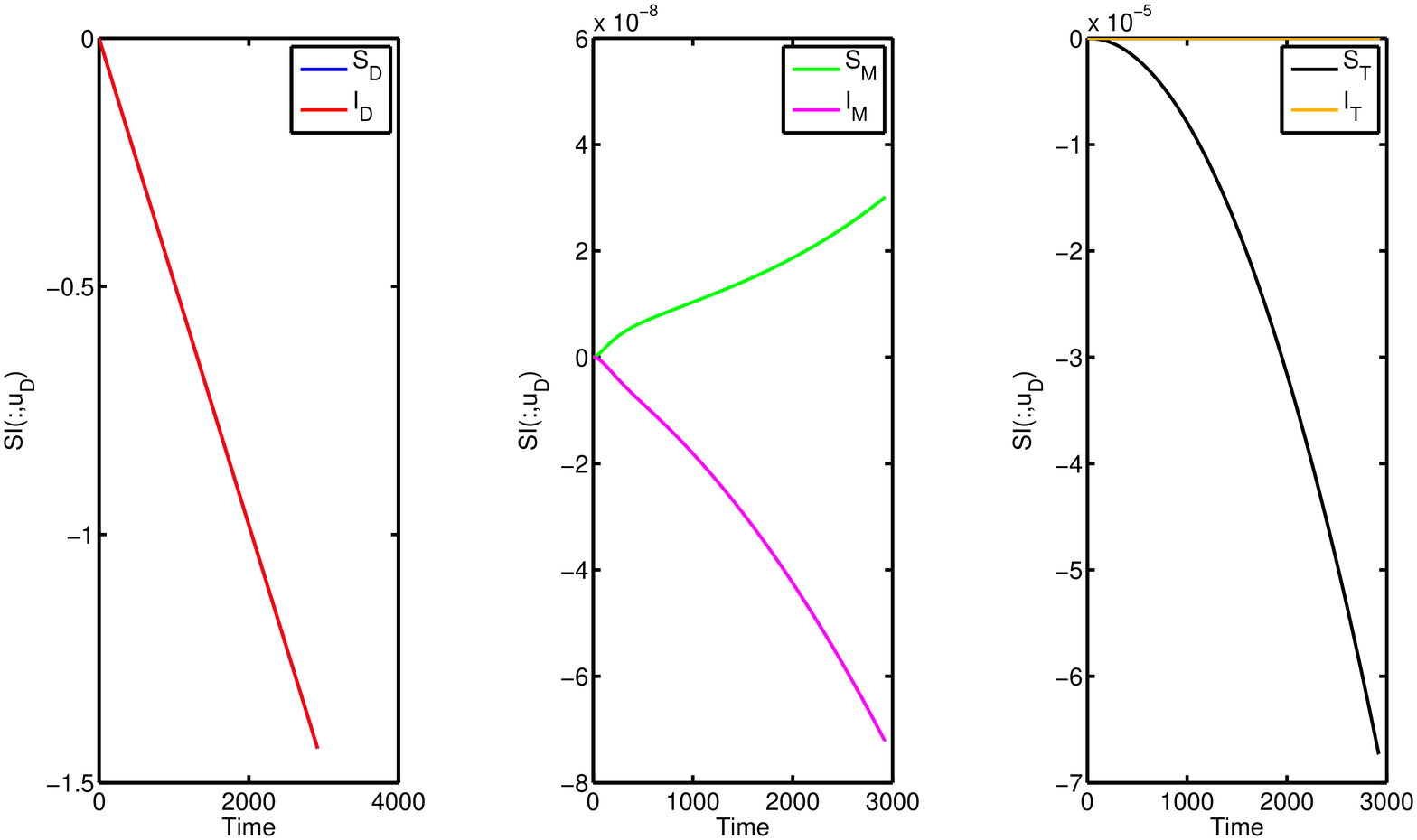}
\end{center}
\caption{\label{fig:7} Population Dynamics (top) and Sensitivity indices w.r.t \hspace{1mm} $\mu_D$ for model \eqref{model1}. Parameters were chosen such that $R_0 > 1$. Initial values are same as in table 1.}
\end{figure}

\begin{figure}[hc!]
\begin{center}
\includegraphics[width=10cm,height=5cm]{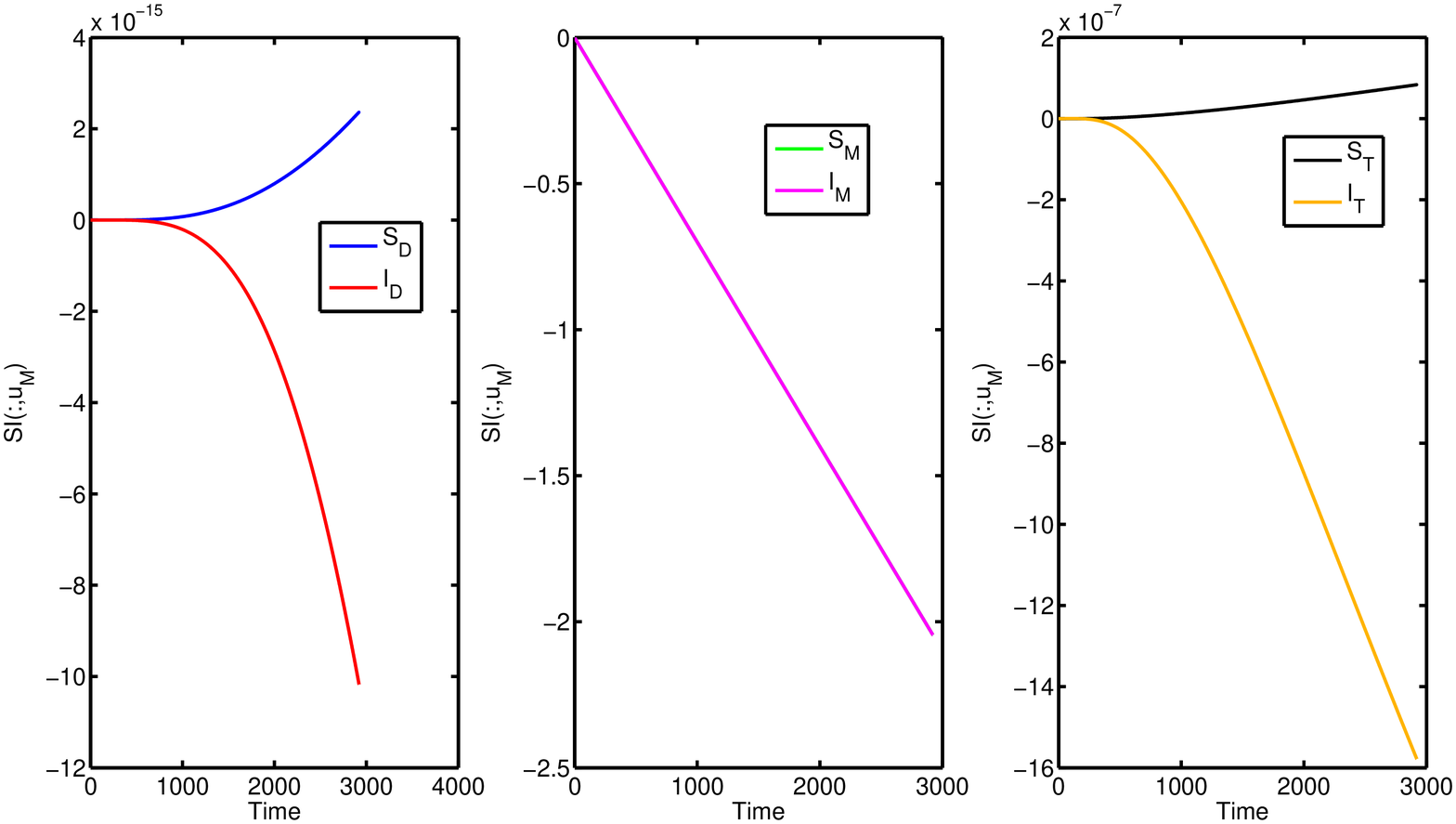}
\includegraphics[width=10cm,height=5cm]{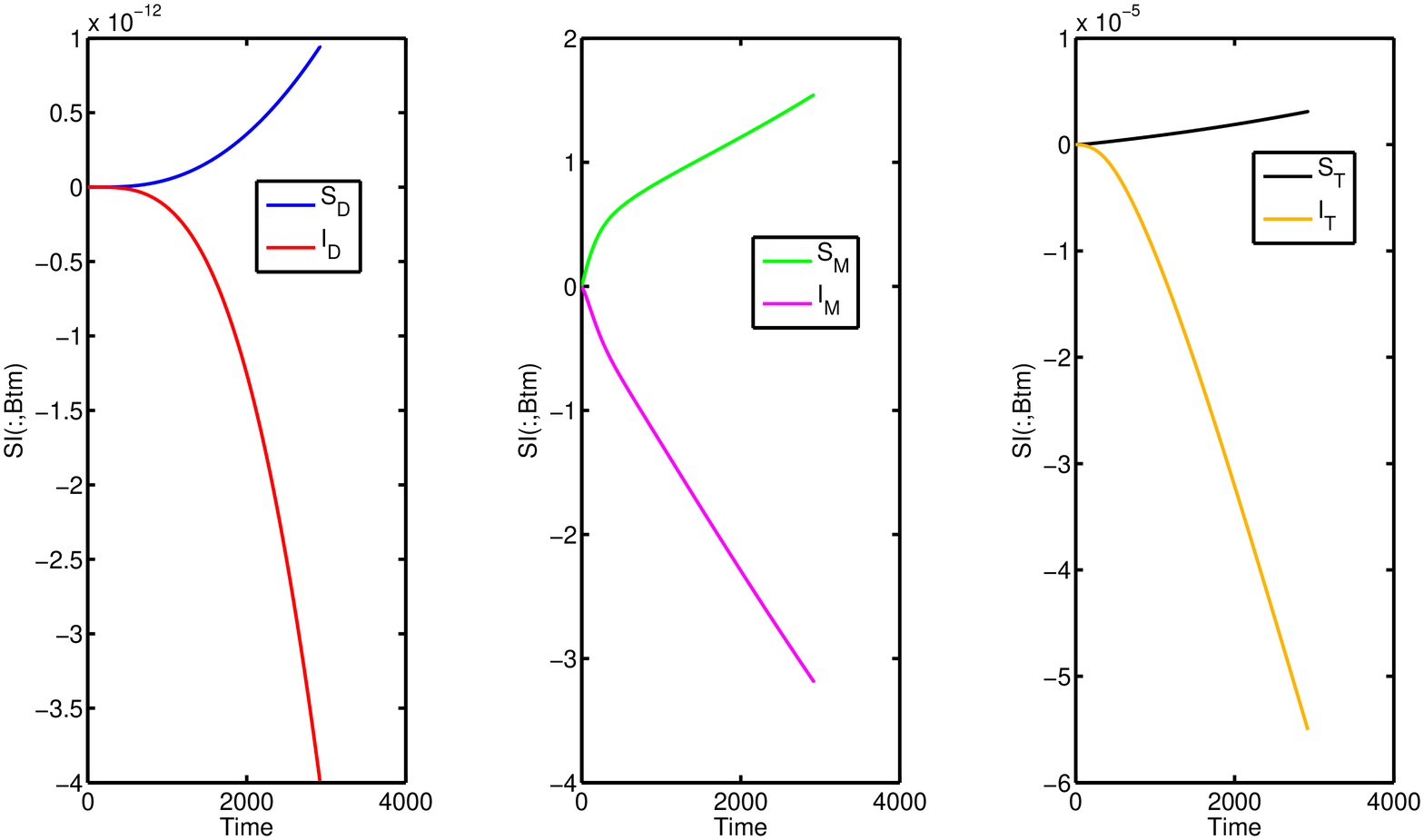}
\includegraphics[width=10cm,height=5cm]{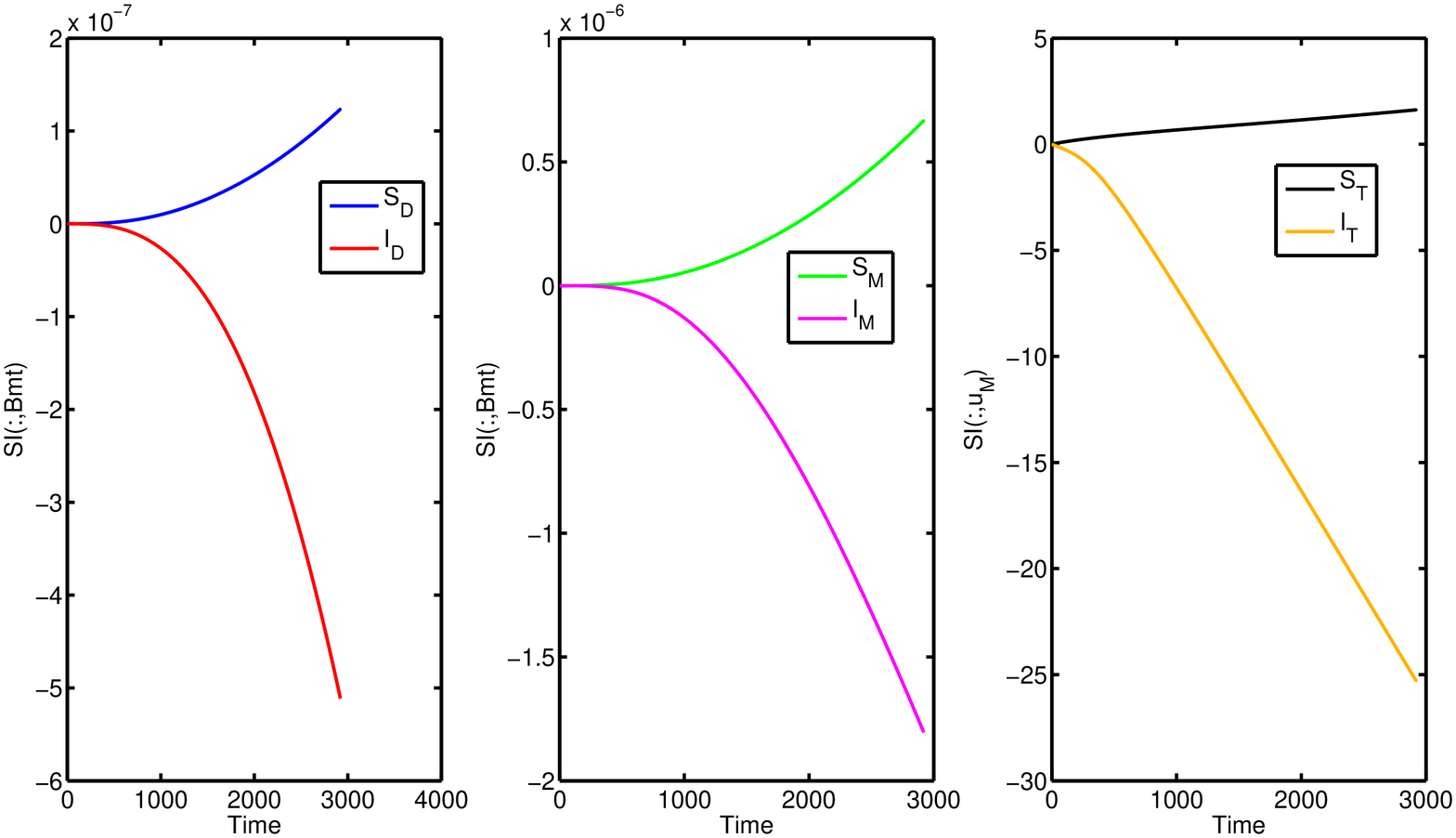}
\end{center}
\caption{\label{fig:8}Sensitivity indices w.r.t \hspace{1mm} $\mu_M$ (top),  $\beta_{TM}$ (middle), and   $\beta_{MT}$ (bottom) for model \eqref{model2}. Parameters were chosen such that $R_0 > 1$. Initial values are same as in table 1.}
\end{figure}

%\begin{figure}[h!]
%\begin{center}
%\includegraphics[width=12cm,height=5cm]{Pop_dynamics_ee_stable.eps}
%\includegraphics[width=12cm,height=5cm]{Sens_dynamics_ee_stable_mud.eps}
%\includegraphics[width=12cm,height=5cm]{Sens_dynamics_ee_stable_mum.eps}
%\end{center}
%\caption{\label{fig:2} Population Dynamics (top) and Sensitivity indices w.r.t \hspace{1mm} $\mu_D$ (middle) and $\mu_M$ (bottom). Parameters were chosen such that $R_0 > 1$ and $R_0 > R_1$. Initial values are same as in table 1.}
%\end{figure}

%\begin{figure}[h!]
%\begin{center}
%\includegraphics[width=12cm,height=5cm]{Pop_dynamics_ee_unstable.eps}
%\includegraphics[width=12cm,height=5cm]{Sens_dynamics_ee_unstable_mud.eps}
%\includegraphics[width=12cm,height=5cm]{Sens_dynamics_ee_unstable_mum.eps}
%\end{center}
%\caption{\label{fig:2} Population Dynamics (top) and Sensitivity indices w.r.t \hspace{1mm} $\mu_D$ (middle) and $\mu_M$ (bottom). Parameters were chosen such that $R_0 < 1$. Initial values are same as in table 1.}
%\end{figure}

%%------------------------------------------------------------------------%%
%%-----------LOGISTIC GROWTH MODIFICATION-----------%%
%%------------------------------------------------------------------------%%

\newpage
\section{Logistic Growth Modification}

We now explore the effects of modeling the Deer population dynamics as logistic model. We consider this modification since large mammal density may be 1-2 orders of magnitudes lower than small mammal host \cite{kelker1947computing}. Deer-specific resource limitations may play an important role in the disease dynamics via the introduction of new susceptible Ticks.

\begin{equation}
\label{model2}
{} \quad
\begin{cases}
\dot{S_{D}} = \Lambda _{D}D \left(1- \frac{D}{K} \right)  -\beta_{TD}\dfrac{I_{T}}{T}S_{D}-\mu _{D}S_{D}\vspace{1.5mm},\\

\dot{I_{D}} = \beta_{TD}\dfrac{I_{T}}{T}S_{D}-\mu _{D}I_{D}\vspace{1.5mm},\\

\dot{S_{M}} = \Lambda _{M}-\beta _{TM}\dfrac {I_{T}} {T}S_{M}-\mu _{M}S_{M} \vspace{1.5mm}\\

\dot{I_{M}} = \beta _{TM}\dfrac {I_{T}} {T}S_{M}-\mu _{M}I_{M} \vspace{1.5mm}\\

\dot{S_{T}} = \Lambda _{T}D-\beta _{MT}S_{T}\dfrac {I_{M}} {M}-\mu_{T}S_{T} \vspace{1.5mm}\\

\dot{I_{T}} = \beta _{MT}S_{T}\dfrac {I_{M}} {M}-\mu_{T}I_{T} \vspace{1.5mm}\\
\end{cases}
\end{equation}

 The modified model keeps all previously held assumptions, except $\Lambda_D$ now describes an intrinsic growth rate for the Deer population and $K$ which denotes carrying capacity.

\subsection{Disease-free and endemic equilibria and their stability}
\begin{theorem} If $\mathcal R_0<1$, then the disease free equilibrium, $ E_0$, is locally asymptotically stable for the system \eqref{model2}.  If $\mathcal R_0>1$, then $E_0$ is unstable. \cite{van2002reproduction}
\end{theorem}
The system \eqref{model2} has a disease free equilibrium denoted by $E_0$, where
$$ E_0=\left\{S_D\to\frac{K(\Lambda _D-\mu_D)}{\Lambda _D},\text{I}_D\to 0, S_M\to \frac{\Lambda _M}{\mu
   _M},\text{I}_M\to 0,S_T\to \frac{K(\Lambda _D-\mu_D) \Lambda _T}{\Lambda_D \mu _T},\text{I}_T\to 0\right\}$$

\noindent Thus, we require $\Lambda_D > \mu_D$ for existence. For stability, evaluating the Jacobian at the equilibria yields the following eigenvalues:

\begin{equation*}
\lambda_{1,2,3,4}=
\left(
\begin{array}{c}
-\mu_{D}\\
-\mu_{T}\\
-\mu_{M}\\
-\Lambda_D + \mu_D\\
\end{array}
\right)
\end{equation*}
\bigskip
\begin{equation*}
\lambda_{5,6} = - \frac{1}{2}\left(\mu_M + \mu_T \pm \sqrt{\mu_M ^2 - 2\mu_T\mu_M + \mu_T ^2 + 4\beta_{MT}\beta_{TM}} \right)
\end{equation*}
Notice that:
\begin{equation*}
\begin{aligned}
A & =(\mu_M + \mu_T)^2 - \left(\mu_M ^2 - 2\mu_T\mu_M + \mu_T ^2 + 4\beta_{MT}\beta_{TM}\right)\\
 &=  4 (-\beta_{MT}\beta_{TM} + \mu_T \mu_M)
\end{aligned}
\end{equation*}

Thus for stability, we additionally require:
\begin{equation*}
\mathcal R_0 = \sqrt{\frac{\beta_{MT}\beta_{TM}}{ \mu_T \mu_M}} < 1
\end{equation*}

\begin{theorem} 
The system \eqref{model2} has a unique endemic equilibrium, $E^*=(S^{*}_{D}, I^{*}_{D}, S^{*}_{M}, I^{*}_{M},S^{*}_{T}, I^{*}_{T})$, iff $\mathcal R_0>1$. The endemic equilibrium $E^*$ is locally asymptotically stable whenever it exists.
\end{theorem}
\bigskip
The endemic equilibria of the model is the following:
$$
S^*_D=\frac{(\Lambda _D-\mu_D) (\mu_T + \beta_{MT}) K \beta_{TM} \mu_D}{((\beta_{MT} \beta_{TM} - \mu_T \mu_M) \beta_{TD} + \beta_{TM} \mu_D (\mu_T + \beta_{MT}))\Lambda_D}
$$

$$
I^*_D=\frac{(\Lambda _D-\mu_D)(\beta_{MT} \beta_{TM} - \mu_T \mu_M) K \beta_{TD}}{((\beta_{MT} \beta_{TM} - \mu_T \mu_M) \beta_{TD} + \beta_{TM} \mu_D (\mu_T + \beta_{MT}))\Lambda_D}
$$

$$
S^*_M= \frac{\Lambda_M (\mu_T + \beta_{MT})}{\beta_{MT} (\mu_M + \beta_{TM})}
$$

$$
I^*_M= \frac{\Lambda_M (\beta_{MT} \beta_{TM} - \mu_T \mu_M)}{\beta_{MT} \mu_M(\mu_M + \beta_{TM})}
$$

$$
S^*_T= \frac{\Lambda_T  (\mu_M + \beta_{TM}) (\Lambda_D - \mu_D) K}{\beta_{TM} \Lambda_D(\mu_T + \beta_{MT})}
$$

$$
I^*_T= \frac{\Lambda_T (\beta_{MT} \beta_{TM} - \mu_T \mu_M) (\Lambda_D - \mu_D) K}{\beta_{TM} \Lambda_D \mu_T (\mu_T + \beta_{MT})}
$$

To verify stability we examine the eigenvalues of the Jacobian matrix at the endemic equilibria:

\begin{equation*}
\lambda_{1,2,3}=
\left(
\begin{array}{c}
-\mu_{T}\\
-\mu_{M}\\
-\Lambda_D +\mu_{M}\\
\end{array}
\right)
\end{equation*}
\begin{equation*}
\lambda_4= - \frac{(\beta_{MT} \beta_{TM} - \mu_T \mu_M) \beta_{TD} + \beta_{TM} \mu_D \mu_T + \mu_D \beta_{TM} \beta_{MT}}{\beta_{TM} (\mu_T + \beta_{MT})}
\end{equation*}
\begin{equation*}
\lambda_{5,6}  = -\left(A_1 \pm \sqrt{A_2}\right)
\end{equation*}
%\newpage
Where: 
\small{\begin{multline*}
A_1 =  \beta_{MT} ^2 \beta_{TM} - \beta_{TM} \mu_T ^2 + \beta_{MT} (- (\beta_{TM} + \mu_M)^2 - 2\beta_{TM}\mu_T) \\   
A_2 = -4(\beta_{TM} + \mu_M )^2 (\beta_{MT} + \mu_M)^2 (\beta_{MT} \beta_{TM} - \mu_T \mu_M) + \beta_{MT}^2\beta_{TM} + \beta_{TM} \mu_T ^2  + \beta_{MT} ((\beta_{TM} + \mu_M)^2 + 2\beta_{TM}\mu_T)^2  
\end{multline*}}

\newpage
$\lambda_4$ is clearly negative whenever $\mathcal R_0 > 1$. Furthermore, simple calculations show that $\lambda_{5,6}$ are also negative since:
\begin{equation*}
A_1 ^2 - A_2 = 4(\mu_M+\beta_{TM})^2 (\mu_T + \beta_{MT})^2 (\beta_{MT}\beta_{TM} - \mu_T \mu_M) > 0
\end{equation*}
when
\begin{equation*}
\mathcal R_0= \sqrt{\frac{\beta_{MT}\beta_{TM}}{ \mu_T \mu_M}} > 1
\end{equation*}

\subsection{Forward Sensitivity Analysis}

The Forward sensitivity equations for model \eqref{model2} are identical to those for model \eqref{model1}. For instance, the equations w.r.t. $\beta_{TM}$ are:\

%\begin{center}
%\begin{}
\[
\frac{\partial S_D}{\partial{\beta_{TM}}}  = \Lambda_D \left(\frac{\partial S_D}{\partial{\beta_{TM}}} + \frac{\partial I_D}{\partial{\beta_{TM}}}\right) \left( 1- \frac{2 D}{K} \right) -\frac{\beta_{TD}}{T^2} \bigg( (I_T \frac{\partial S_D}{\partial{\beta_{TM}}}\
  \frac{\partial I_T}{\partial{\beta_{TM}}} )T - I_TS_D (\frac{\partial S_T}{\partial{\beta_{TM}}} + \frac{\partial I_T}{\partial{\beta_{TM}}})\bigg) - \beta_{TM} \frac{\partial S_D}{\partial {\beta_{TM}}} - S_D\]
   \[\frac{\partial I_D}{\partial {\beta_{TM}}}  =  \frac{\beta_{TD}}{T^2} \bigg( (I_T \frac{\partial S_D}{\partial{\beta_{TM}}} + S_D \frac{\partial I_T}{\partial{\beta_{TM}}} )T - I_TS_D (\frac{\partial S_T}{\partial{\beta_{TM}}} + \frac{\partial I_T}{\partial{\beta_{TM}}})\bigg) - \beta_{TM} \frac{\partial I_D}{\partial {\beta_{TM}}}\]
  \[\frac{\partial S_M}{\partial {\beta_{TM}}}  =  -\frac{\beta_{TM}}{T^2} \bigg( (I_T \frac{\partial S_M}{\partial{\beta_{TM}}} + S_M \frac{\partial I_T}{\partial{\beta_{TM}}} )T - I_TS_M (\frac{\partial S_T}{\partial{\beta_{TM}}} + \frac{\partial I_T}{\partial{\beta_{TM}}})\bigg) - \mu_M \frac{\partial S_M}{\partial {\beta_{TM}}} - S_M\frac{I_T}{T}\]
  \[\frac{\partial I_M}{\partial {\beta_{TM}}}  = \frac{\beta_{TM}}{T^2} \bigg( (I_T \frac{\partial S_M}{\partial{\beta_{TM}}} + S_M \frac{\partial I_T}{\partial{\beta_{TM}}} )T - I_TS_M (\frac{\partial S_T}{\partial{\beta_{TM}}} + \frac{\partial I_T}{\partial{\beta_{TM}}})\bigg) - \mu_M \frac{\partial I_M}{\partial {\beta_{TM}}} + S_M\frac{I_T}{T}\]
  \[\frac{\partial S_T}{\partial {\beta_{TM}}}  = \Lambda_T \bigg(\frac{\partial S_D}{\partial {\beta_{TM}}} + \frac{\partial I_D}{\partial {\beta_{TM}}}\bigg) - \frac{\beta_{MT}}{M^2} \bigg( (S_T \frac{\partial I_M}{\partial{\beta_{TM}}} + I_M \frac{\partial S_T}{\partial{\beta_{TM}}} )M - S_TI_M (\frac{\partial S_M}{\partial{\beta_{TM}}} + \frac{\partial I_M}{\partial{\beta_{TM}}})\bigg) - \mu_T \frac{\partial S_T}{\partial {\beta_{TM}}}\]
  \[\frac{\partial I_T}{\partial {\beta_{TM}}}  =  \frac{\beta_{MT}}{M^2} \bigg( (S_T \frac{\partial I_M}{\partial{\beta_{TM}}} + I_M \frac{\partial S_T}{\partial{\beta_{TM}}} )M - S_TI_M (\frac{\partial S_M}{\partial{\beta_{TM}}} + \frac{\partial I_M}{\partial{\beta_{TM}}})\bigg) - \mu_T \frac{\partial I_T}{\partial {\beta_{TM}}}
\]
%\end{}
%\end{center}

\newpage
The results show some interesting contrasts with model \eqref{model1}. The sensitivity of the susceptible Deer population to $\mu_D$ rebounds once the population reaches carrying capacity, while the infected Deer class remains relatively more sensitive. The sensitivity of the infected mice class to $\mu_D$ is never positive (compared to model \eqref{model1})  indicating that deer harvesting always decreases overall infection.

\begin{figure}[hc!]
\begin{center}
\includegraphics[width=10cm,height=5cm]{Pop_dynamics_ee_stable_log.eps}
\includegraphics[width=10cm,height=5cm]{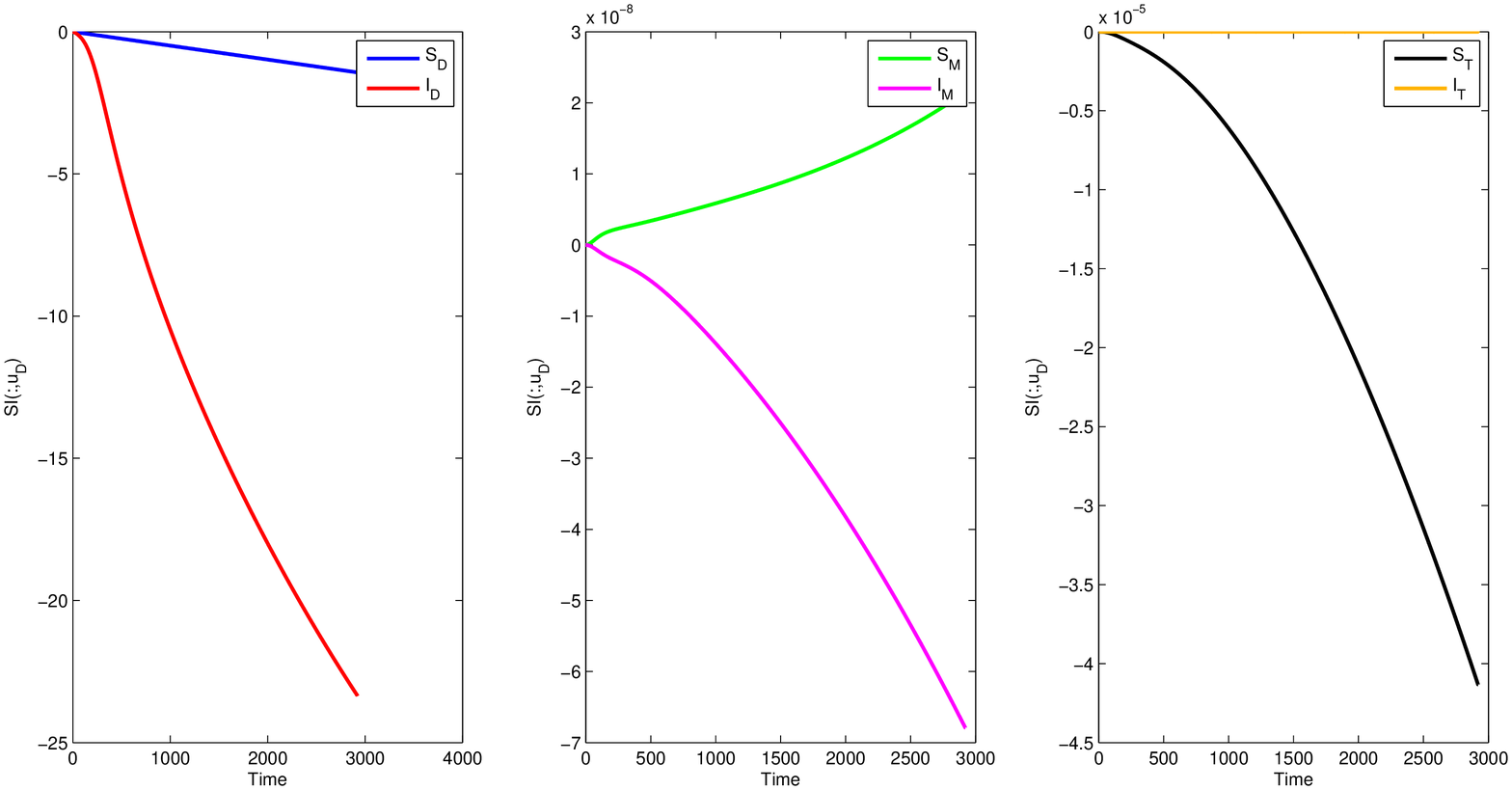}
\end{center}
\caption{\label{fig:7} Population Dynamics (top) and Sensitivity indices w.r.t \hspace{1mm} $\mu_D$ for model \eqref{model2}. Parameters were chosen such that $R_0 > 1$. K = 5000, and initial values are same as in table 1.}
\end{figure}

\begin{figure}[hc!]
\begin{center}
\includegraphics[width=10cm,height=5cm]{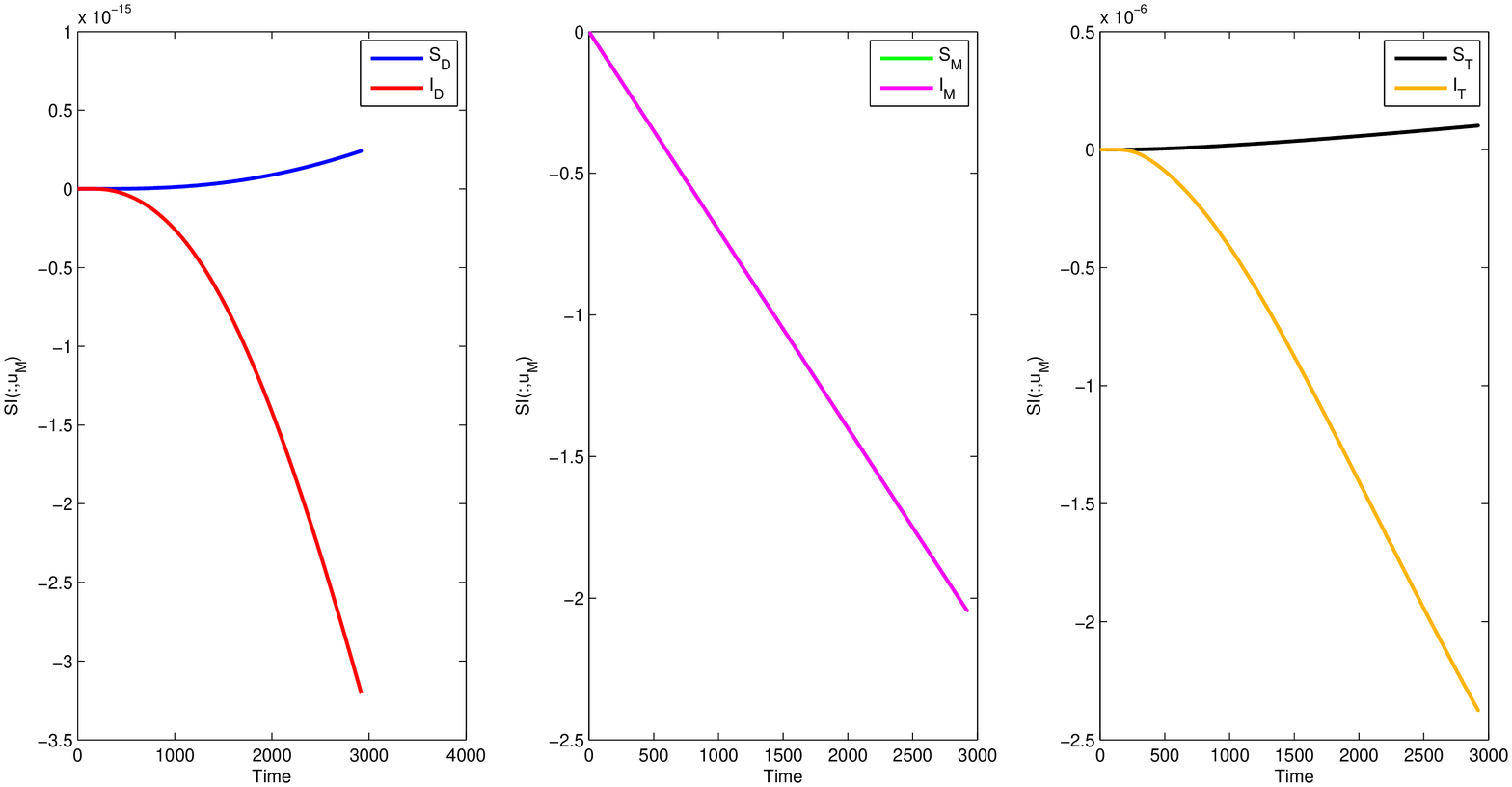}
\includegraphics[width=10cm,height=5cm]{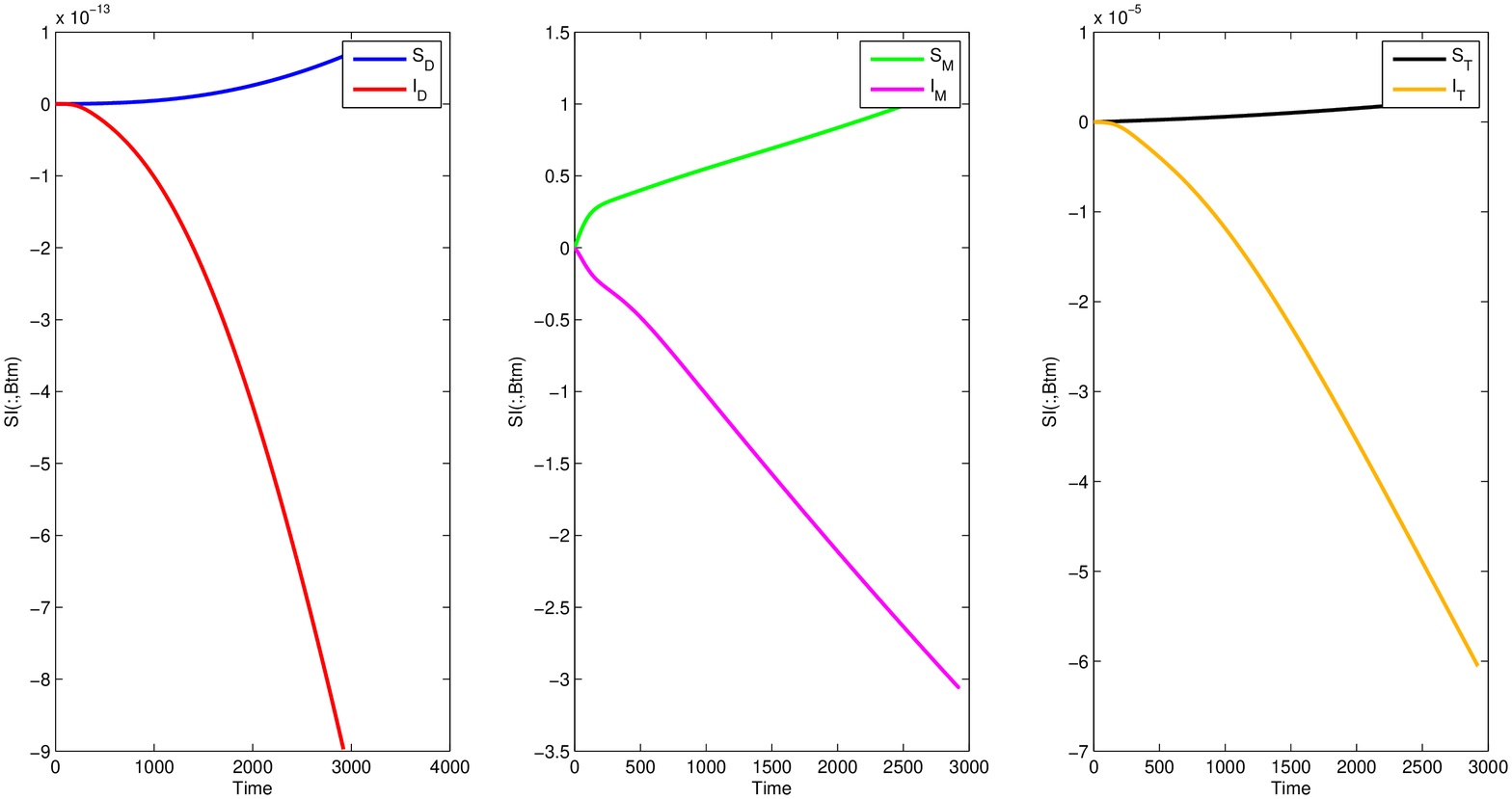}
\includegraphics[width=10cm,height=5cm]{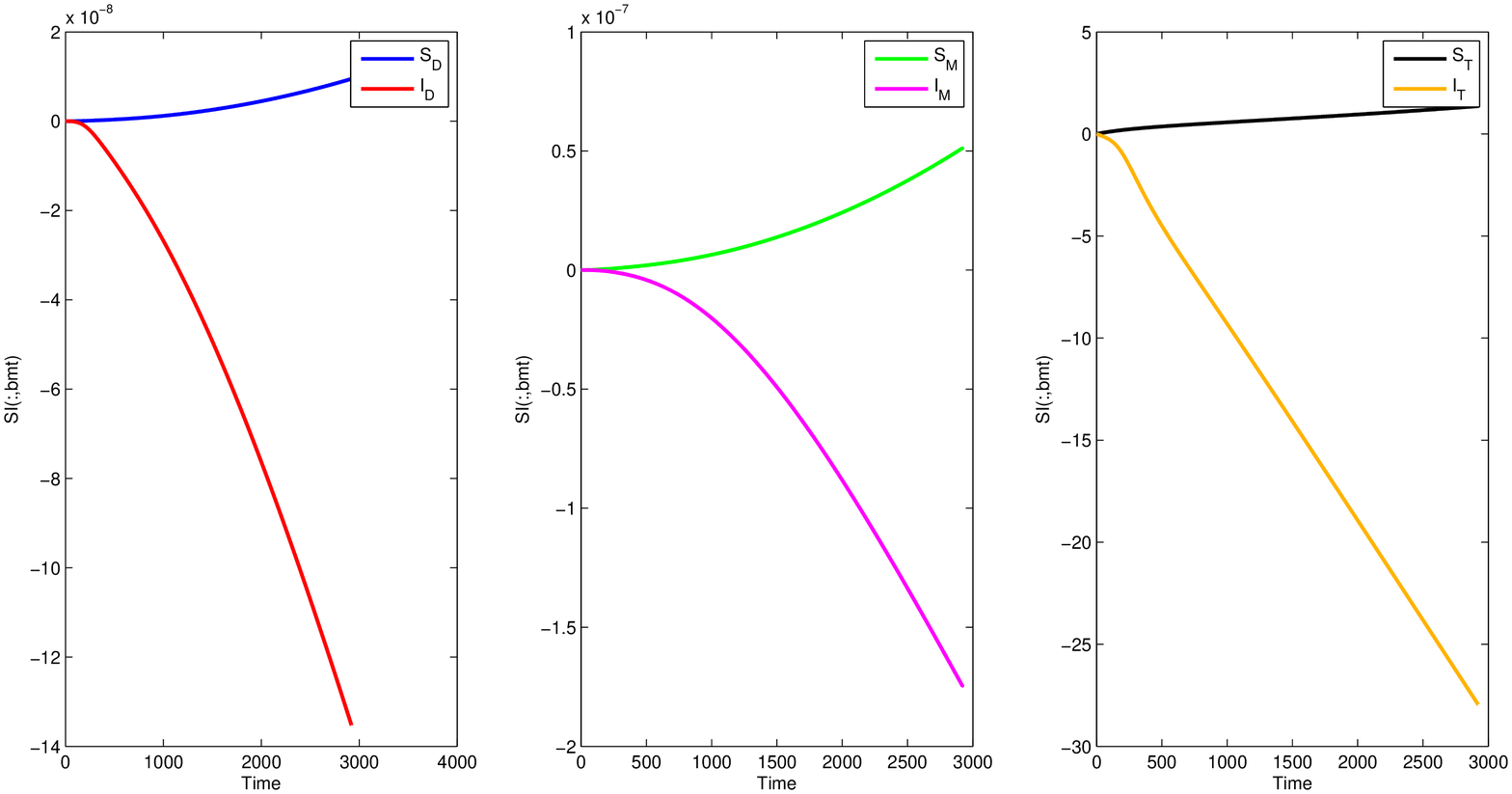}
\end{center}
\caption{\label{fig:8}Sensitivity indices w.r.t \hspace{1mm} $\mu_M$ (top),  $\beta_{TM}$ (middle), and   $\beta_{MT}$ (bottom) for model \eqref{model2}. Parameters were chosen such that $R_0 > 1$. K = 5000, and initial values are same as in table 1.}
\end{figure}

%\begin{figure}[hc!]
%\begin{center}
%\includegraphics[width=12cm,height=5cm]{Pop_dynamics_ee_unstable_log.eps}
%\includegraphics[width=12cm,height=5cm]{Sens_dynamics_ee_unstable_mu_d_log.eps}
%\includegraphics[width=12cm,height=5cm]{Sens_dynamics_ee_unstable_mu_m_log.eps}
%\end{center}
%\caption{\label{fig:5} Population Dynamics (top) and Sensitivity indices w.r.t \hspace{1mm} $\mu_D$ (middle) and $\mu_M$ (bottom) for model \eqref{model2}. Parameters were chosen such that $R_0 < 1$. K= 5000 and initial values are same as in table 1.}
%\end{figure}

\newpage
             û
\newpage
\bibliographystyle{apalike}

%%------------------------------------------------------------------------%%
%%-------------------------DISCUSSION----------------------------%%
%%------------------------------------------------------------------------%%

\section{Discussion}

The rapid increase in Lyme disease cases in humans highlights the need to control the spread of infected black-legged ticks. We present a compartmental, SI model of the spread of Lyme disease for deer, tick and mice compartments in which the populations are grouped in susceptible and infected classes.  The host, reservoir and vector dynamics in this particular model is crucial due to the life cycle of the tick. They reproduce mostly on deer, but get infected with the bacteria when feeding on mice during the nymphal stage. This implies that the birthrate of the ticks is dependent on the density of the deer. 
The parameters chosen to perform sensitivity analysis are the death rates of the reservoir host and the vector, as well as transmission rates between the two. 
The possibility of altering these parameters by introducing control methods in a real population of deer, mice and ticks prompted us to focus on said death rates and transmission rates. Analyzing the death rates of ticks, for example, would be the most effective but not realistic in terms of implementing policies. The dynamic sensitivity analysis of this model was performed to observe how the sensitivity changes over time when parameters relevant for control methods are perturbed. \\

The evolution of the susceptible and infected classes over a period of 3000 days indicates a pronounced increase in the numbers of susceptible mice which decreases after day 2000. This is mainly due to the values of the per capita death rate of the mice as well as the slow increase of the infected mice. The population of the deer remains relatively constant in number over the time interval of the simulation. Dynamic and static sensitivity analyses were performed on the model at the reproductive number and at the endemic equilibrium respectively. The static sensitivity of the reproductive number with respect to the death rates of the deer and mice yielded negative sensitivity values, which implies a reduction of the basic reproduction number. As $\mathcal R_0$ depends on transmission rates and death rates of mice and ticks, increasing the death rate of the deer or mice will cause this number to reduce, as $\mu_M$ is in the denominator of the expression. In the case of the deer, as we are reducing the reproductive host of the ticks, the population of ticks declines and this change is reflected in the decrease of the reproductive number \cite{diuk2006spatiotemporal}. Decreasing the transmission rates between mice and ticks produces a negative sensitivity for the $R_0$. This can be intuitively seen as the transmission rates are located in the denominator of $ R_0$; it follows that any increase in the transmission rates would result in a decrease in $R_0$. At the endemic equilibrium, the death rate of the deer had a larger impact on the infected tick population compared to the effect it had on the infected mice population. The sensitivity index at the endemic equilibrium of the infected mice population with respect to the death rate of the deer was zero; in spite of this, we cannot conclude that there is no effect but rather that it is a secondary effect that is not reflected in this first order approximation. Decreasing the population of mice has a large impact on the infected tick population, as the reservoir host of the bacteria is reduced thus infecting less ticks. It had a small negative impact on the infected deer population, which was an indirect consequence of reducing the infected tick population. \\

The dynamical sensitivity analysis with respect to deer death rate reflected a direct negative impact on the infected deer population. The population of susceptible ticks decreases as the host on which they reproduce is culled. As a consequence there is a small increase in the susceptible mice population and a decline in infected mice.  Lowering the number of deer reduces the population of ticks quite efficiently. However, the increase of infected mice is a problem because the number of infected ticks will start increasing over time as the susceptible mice decrease. Decreasing the numbers of both susceptible and infected mice constantly in an interval of time impacts the tick population directly. The effect is most pronounced in the infected class which experiences a dramatic reduction. Infected mice decline, causing an important reduction in the infected tick population, which is a direct consequence of the transmission rate of the disease from mice to ticks \cite{bosler1984prevalence}. Additionally, as mice act as the reservoir host for Lyme disease \cite{control1995}, when the number of infected mice decreases, so does the infected tick population. Although the tick population does not drastically decrease, the importance of increasing the mouse death rate is that we are reducing the pool of infectious hosts that spread the disease. Increasing the death rate of the mice has some benefits over increasing the death rate of the deer, in spite of the small impact on the infected tick population.
Ticks have preferred hosts, however their survival is a direct result of their adaptability \cite{ostfeld2010lyme}. Thus, if there is a reduction of the deer, the reduced tick population could potentially switch hosts, increasing the number of infected mice \cite{barbour1993biological}.\\

Altering the transmission rates is also a possibility when it comes to trying to control Lyme disease. The transmission rates between mice and ticks depend on the probability of contact and biting rate. The strategy used by the CDC in a study currently underway in Connecticut was to place bait boxes with food for mice. These boxes also include a wick with the pesticide, fipronil, which kills tick on the mice without harming the mammal \cite{Lymedata}. In the dynamic sensitivity analysis performed with respect to the transmission rates we obtained different sensitivities indices depending on whether the transmission rate was from tick to deer or vice versa. When decreasing $\beta_{MT}$, the transmission rate from mice to tick, there is a considerable decrease in the infected tick population, but a very small impact on the infected mice population which is a secondary effect. Decreasing the transmission rate from tick to mice impacts the infected mice population negatively, but has a small effect in the decrease of infected tick populations.

\subsection{Future Work}

This model has the limitation of not including the life cycle of \emph{I. scapularis} and the different hosts that it inhabits in each stage of its life. This would be relevant to introduce control measures in a specific host at a particular life stage of the tick to prevent the tick from acquiring the bacteria (rodent control) or reducing the total population of ticks (large mammal control). Another possible issue with the model is that our control measures consist of modifying death rates constantly over time; this is not realistic in the sense that controlling deer would be done once a year, during the hunting season, which is a yearly occurrence in a very short amount of time. For future work, we can consider introducing a harvesting term on the deer that reflects the hunting season more accurately and/or introducing the different life stages of the tick. We would also like to introduce a two-patch model that would incorporate migration, geographical restrictions, and variable harvesting restrictions per population. Including seasonality into the model would also be an interesting avenue for future study to increase accuracy. Additionally, we would like to calculate the total sensitivity of our populations as opposed to the instantaneous dynamic sensitivity. For example, from the functional $J(I_D) = \int^T_{t=0} \frac{dI_D}{dt}\,dt$, we would use the adjoint method to determine the sensitivity of the total $I_D$ population.

%%-------------------------------------------------------------------------------------%%
%%-------------------------ACKNOWLEDGMENTS----------------------------%%
%%-------------------------------------------------------------------------------------%%

\section*{Acknowledgments} 
We would like to thank Dr.~Carlos Castillo-Chavez, Executive Director of the Mathematical and Theoretical Biology Institute (MTBI), for giving us the opportunity to participate in this research program.  We would also like to thank Co-Executive Summer Directors Dr.~Erika T.~Camacho and Dr.~Stephen Wirkus for their efforts in planning and executing the day to day activities of MTBI. We also want to give special thanks to Xiaoguang Zhang for his help with the stability analysis. This research was conducted in MTBI at the Mathematical, Computational and Modeling Sciences Center (MCMSC) at Arizona State University (ASU). This project has been partially supported by grants from the National Science Foundation (NSF - Grant DMPS-1263374), the National Security Agency (NSA - Grant H98230-13-1-0261), the Office of the President of ASU, and the Office of the Provost of ASU.

%%------------------------------------------------------------------------%%
%%-----------------------BIBLIOGRAPHY---------------------------%%
%%------------------------------------------------------------------------%%
\bibliographystyle{plain}
\bibliography{lyme}{}

%%-----------------------------------------------------------------------%%
%%--------------------------APPENDIX------------------------------%%
%%-----------------------------------------------------------------------%%
\newpage
\section{Appendix}
\begin{table}[h!]\caption{Sources of the variables in the modeling 
framework}
\centering 
\begin{tabularx}{\textwidth}{>{} l|X|c|}
\toprule
Parameter & Source \\ [0.5ex]
\toprule
$D$ & \cite{roseberry1998habitat}\\ [0.5ex]
$T$ & \cite{diuk2006spatiotemporal}\\[0.5 ex]
$S_{T}$ & \cite{diuk2006spatiotemporal}\\ [0.5ex]	
$I_{T}$ & \cite{diuk2006spatiotemporal}\\ [0.5ex]
$\Lambda_{D}$ &   \cite{kelker1947computing}\\[0.5 ex] 
$\Lambda_{M}$ &  \cite{fahrig1985habitat}\cite{jacquot1998recruitment}\\[0.5ex]
$\Lambda_{T}$ & \cite{levi2012deer}  \\[0.5ex]
$\beta_{TD}$ & \cite{levi2012deer}	\\[0.5ex]
$\beta_{TM}$  & \cite{levi2012deer}  \\[0.5 ex]
$\beta_{MT}$ & \cite{levi2012deer}  \\[0.5 ex]
$\mu_{D}$  & \cite{nelson1986mortality} \\[0.5 ex]
$\mu_{M}$ & \cite{fahrig1985habitat}\cite{ruan1999differential}\\[0.5 ex]
$\mu_{T}$ &\cite{levi2012deer} \\[0.5 ex]
\bottomrule
\end{tabularx}
\end{table}

%Any Appendix goes here as Appendix A, Appendix B, etc. as separate sections.
%
\end{document}